\documentclass[11pt]{elsarticle}

\usepackage{amsmath}
\usepackage{amssymb}
\usepackage{subcaption}
\usepackage{url}
\usepackage[top=5em, bottom=5em, left=3.5em, right=3.5em]{geometry}
\usepackage{tabularx}
\usepackage{supertabular}
\usepackage{multirow}
\usepackage{boxedminipage}
\usepackage{eurosym}
\usepackage{xcolor}
\usepackage{mathtools}
\usepackage{bm}
\usepackage{comment}
\usepackage{float}
\usepackage{graphicx}
\usepackage{enumerate}

\newtheorem{remark}{Remark}

\begin{document}
	
\newcommand{\add}[1]{\textcolor{black}{#1}}

\newcommand{\del}[1]{\textcolor{red}{\st{#1}}}
	
% Sets
\newcommand{\setDemNode}{\ensuremath{\Omega^D_{t,n}}}
\newcommand{\setDemAll}{\ensuremath{\Omega^D_{t}}}
\newcommand{\setGenNode}{\ensuremath{\Omega^G_{t,n}}}
\newcommand{\setGenAll}{\ensuremath{\Omega^G_{t}}}
\newcommand{\setNodesPlus}{\ensuremath{\mathcal{N^+}}}
\newcommand{\setNodes}{\ensuremath{\mathcal{N}}}
\newcommand{\setLines}{\ensuremath{\mathcal{L}}}
\newcommand{\setTime}{\ensuremath{\mathcal{T}}}
\newcommand{\setLumpy}{\ensuremath{\mathcal{M}}}

% shorthand
\newcommand{\forallt}{\ensuremath{{\forall t \in \setTime}}}
\newcommand{\forallm}{\ensuremath{{\forall m \in \setLumpy}}}
\newcommand{\foralln}{\ensuremath{\forall n \in \setNodes}}
\newcommand{\forallnplus}{\ensuremath{{\forall n \in \setNodesPlus}}}
\newcommand{\forallij}{\ensuremath{{\forall (i,j) \in \setLines}}}

% Variables
\newcommand{\uBinary}{\ensuremath{u_{i,j,m}}}

\newcommand{\dem}{\ensuremath{d^p_{t,n,k}}}
\newcommand{\demReactive}{\ensuremath{d^q_{t,n,k}}}
\newcommand{\gen}{\ensuremath{g^p_{t,n,k}}}
\newcommand{\genReactive}{\ensuremath{g^q_{t,n,k}}}

\newcommand{\wii}{\ensuremath{W}^{ii}_{t,i}}
\newcommand{\wzerozero}{\ensuremath{W}^{ii}_{t,0}}
\newcommand{\wjj}{\ensuremath{W}^{ii}_{t,j}}

\newcommand{\wijP}{\ensuremath{W}^{ij,p}_{t,i,j}}
\newcommand{\wijQ}{\ensuremath{W}^{ij,q}_{t,i,j}}

\newcommand{\wjiP}{\ensuremath{W}^{ji,p}_{t,j,i}}
\newcommand{\wjiQ}{\ensuremath{W}^{ji,q}_{t,j,i}}

\newcommand{\admitance}{\ensuremath{y}_{i,j,m}}
\newcommand{\aRealAdmittance}{\ensuremath{a}_{i,j,m}}
\newcommand{\bImagAdmittance}{\ensuremath{e}_{i,j,m}}

\newcommand{\complexInj}{\ensuremath{p}_{n}}
\newcommand{\activeInj}{\ensuremath{p_{t,n}}}
\newcommand{\activeInjZero}{\ensuremath{p_{t,0}}}
\newcommand{\reactiveInj}{\ensuremath{q_{t,n}}}
\newcommand{\reactiveInjZero}{\ensuremath{q_{t,0}}}

\newcommand{\reserveUp}{\ensuremath{r^{up}_{t,n}}}
\newcommand{\reserveDown}{\ensuremath{r^{down}_{t,n}}}

\newcommand{\tariff}{\ensuremath{\tau}}
\newcommand{\priceUniqueDemFixed}{\ensuremath{\pi^D_\mathcal{A}}}

% Parameters
\newcommand{\demFixed}{\ensuremath{{D}_{t,n}}}
\newcommand{\demMax}{\ensuremath{{d}^{p,max}_{t,n,k}}}
\newcommand{\demMin}{\ensuremath{{d}^{p,min}_{t,n,k}}}

\newcommand{\genMax}{\ensuremath{{g}^{p,max}_{t,n,k}}}
\newcommand{\genMin}{\ensuremath{{g}^{p,min}_{t,n,k}}}

\newcommand{\demReactiveMax}{\ensuremath{{d}^{q,max}_{t,n,k}}}
\newcommand{\demReactiveMin}{\ensuremath{{d}^{q,min}_{t,n,k}}}

\newcommand{\genReactiveMax}{\ensuremath{{g}^{q,max}_{t,n,k}}}
\newcommand{\genReactiveMin}{\ensuremath{{g}^{q,min}_{t,n,k}}}

\newcommand{\Flumpy}{\ensuremath{{F}^{}_{i,j,m}}}
\newcommand{\flowMax}{\ensuremath{{F}^{max}_{i,j}}}
\newcommand{\flowMin}{\ensuremath{{F}^{min}_{i,j}}}

\newcommand{\voltageMagnitudeMax}{\ensuremath{{v}^{max}_{t,i}}}
\newcommand{\voltageMagnitudeMin}{\ensuremath{{v}^{min}_{t,i}}}

\newcommand{\Kvar}{\ensuremath{{K}^{var}_{i,j,m}}}
\newcommand{\Kfix}{\ensuremath{{K}^{fix}_{i,j,m}}}
\newcommand{\Kop}{\ensuremath{{K}^{op}_{}}}
\newcommand{\Ktot}{\ensuremath{{K}^{tot}_{}}}

\newcommand{\priceActivePowerZero}{\ensuremath{{c}^{p}_{t,0}}}
\newcommand{\priceReactivePowerZero}{\ensuremath{{c}^{q}_{t,0}}}
\newcommand{\priceReserveUp}{\ensuremath{{c}^{up}_{t,0}}}
\newcommand{\priceReserveDown}{\ensuremath{{c}^{down}_{t,0}}}
\newcommand{\priceDem}{\ensuremath{{c}^d_{t,n,k}}}
\newcommand{\priceGen}{\ensuremath{{c}^g_{t,n,k}}}

\newcommand{\sensitivityInjection}{}

% duals
\newcommand{\nodalPrice}{\ensuremath{\pi^p_{t,n}}}
\newcommand{\nodalPriceRective}{\ensuremath{\pi^q_{t,n}}}
\newcommand{\lambdaPowerActive}{\ensuremath{\lambda^p_{t,n}}}
\newcommand{\lambdaPowerActiveI}{\ensuremath{\lambda^p_{t,i}}}
\newcommand{\lambdaPowerActiveJ}{\ensuremath{\lambda^p_{t,j}}}
\newcommand{\lambdaPowerReactive}{\ensuremath{\lambda^q_{t,n}}}
\newcommand{\lambdaPowerReactiveI}{\ensuremath{\lambda^q_{t,i}}}
\newcommand{\lambdaPowerReactiveJ}{\ensuremath{\lambda^q_{t,j}}}
\newcommand{\muMax}{\ensuremath{\mu^{max}_{t,i,j}}}
\newcommand{\muMin}{\ensuremath{\mu^{min}_{t,i,j}}}
\newcommand{\phiActiveDemMax}{\ensuremath{\varphi^{d,p,max}_{t,n,k}}}
\newcommand{\phiActiveDemMin}{\ensuremath{\varphi^{d,p,min}_{t,n,k}}}
\newcommand{\phiActiveGenMax}{\ensuremath{\varphi^{g,p,max}_{t,n,k}}}
\newcommand{\phiActiveGenMin}{\ensuremath{\varphi^{g,p,min}_{t,n,k}}}
\newcommand{\phiReactiveDemMax}{\ensuremath{\varphi^{d,q,max}_{t,n,k}}}
\newcommand{\phiReactiveDemMin}{\ensuremath{\varphi^{d,q,min}_{t,n,k}}}
\newcommand{\phiReactiveGenMax}{\ensuremath{\varphi^{g,q,max}_{t,n,k}}}
\newcommand{\phiReactiveGenMin}{\ensuremath{\varphi^{g,q,min}_{t,n,k}}}
\newcommand{\rhoUp}{\ensuremath{\rho^{up}_{t,n}}}
\newcommand{\rhoDown}{\ensuremath{\rho^{down}_{t,n}}}

\newcommand{\chiMax}{\ensuremath{\chi^{max}_{t,n}}}
\newcommand{\chiMin}{\ensuremath{\chi^{min}_{t,n}}}

\newcommand{\etaOne}{\ensuremath{\eta^{a}_{t,i,j}}}
\newcommand{\etaTwo}{\ensuremath{\eta^{b}_{t,i,j}}}
\newcommand{\etaThree}{\ensuremath{\eta^{c}_{t,i,j}}}
\newcommand{\gammaSOCP}{\ensuremath{\gamma^{}_{t,i,j}}}

\newcommand{\epsilonP}{\ensuremath{\epsilon^{p}_{t,i,j}}}
\newcommand{\epsilonQ}{\ensuremath{\epsilon^{q}_{t,i,j}}}

% auxiliary variables
\newcommand{\yWiiu}{\ensuremath{y^{w^{ii},u}_{t,i,j,m}}}
\newcommand{\yWjju}{\ensuremath{y^{w^{ii},u}_{t,j,i,m}}}
\newcommand{\yWijPu}{\ensuremath{y^{w^{ij}p,u}_{t,i,j,m}}}
\newcommand{\yWijQu}{\ensuremath{y^{w^{ij}q,u}_{t,i,j,m}}}
\newcommand{\yWjiPu}{\ensuremath{y^{w^{ji}p,u}_{t,i,j,m}}}
\newcommand{\yWjiQu}{\ensuremath{y^{w^{ji}q,u}_{t,i,j,m}}}
\newcommand{\ylambdaPu}{\ensuremath{y^{\lambda,p,u}_{t,i,j,m}}}
\newcommand{\ylambdaPuJ}{\ensuremath{y^{\lambda,p,u}_{t,j,i,m}}}
\newcommand{\ylambdaQu}{\ensuremath{y^{\lambda,q,u}_{t,i,j,m}}}
\newcommand{\ylambdaQuJ}{\ensuremath{y^{\lambda,q,u}_{t,j,i,m}}}
\newcommand{\ymuMaxU}{\ensuremath{y^{\muMax, u}_{t,i,j,m}}}
\newcommand{\ymuMinU}{\ensuremath{y^{\muMin,u}_{t,i,j,m}}}

\newcommand{\yPU}{\ensuremath{y^{p,u}_{t,i,j,m,k}}}
\newcommand{\yQU}{\ensuremath{y^{q,u}_{t,i,j,m,k}}}
\newcommand{\yBetaU}{\ensuremath{y^{\beta,u}_{t,i,j,m,k}}}

\begin{frontmatter}
	
	\title{Electricity prices and tariffs to keep everyone happy: a framework for fixed and nodal prices coexistence in distribution grids with optimal tariffs for investment cost recovery}
	
	%author names and affiliations
	\author[smith]{Iacopo~Savelli\corref{cor1}}
	\ead{iacopo.savelli@smithschool.ox.ac.uk}
	
	\author[eng]{Thomas~Morstyn}
	\ead{thomas.morstyn@ed.ac.uk}

	\cortext[cor1]{Corresponding author}
	
	\address[smith]{Smith School of Enterprise and the Environment, University of Oxford, South Parks Road, Oxford, UK}
	\address[eng]{School of Engineering, University of Edinburgh, Robert Stevenson Road, Edinburgh, UK}

\begin{abstract}
Some consumers, particularly households, are unwilling to face volatile electricity prices, and they can perceive as unfair price differentiation in the same local area. For these reasons, nodal prices in distribution networks are rarely employed. However, the increasing availability of renewable resources and emerging price-elastic behaviours pave the way for the effective introduction of marginal nodal pricing schemes in distribution networks. The aim of the proposed framework is to show how traditional non-flexible consumers can coexist with flexible users in a local distribution area. Flexible users will pay nodal prices, whereas non-flexible consumers will be charged a fixed price derived from the underlying nodal prices. Moreover, the developed approach shows how a distribution system operator should manage the local grid by optimally determining the lines to be expanded, and the collected network tariff levied on grid users, while accounting for both congestion rent and investment costs. The proposed model is formulated as a non-linear integer bilevel program, which is then recast as an equivalent single optimization problem, by using integer algebra and complementarity relations. The power flows in the distribution area are modelled by resorting to a second-order cone relaxation, whose solution is exact for radial networks under mild assumptions. The final model results in a mixed-integer quadratically constrained program, which can be solved with off-the-shelf solvers. Numerical test cases based on both 5-bus and 33-bus networks are reported to show the effectiveness of the proposed method. 
\end{abstract}


\end{frontmatter}

\onecolumn
\linespread{1.25}

\section{Introduction}\label{sec:introduction}

By 2050, half of all European Union citizens could produce energy from renewable energy resources, with a total electricity generation capable of satisfying up to 45\% of the whole European energy demand \cite{kampman2016potential}. 
In particular, households are expected to become more aware and engaged into the energy system \cite{bosman2012planning}. For example, they could provide services to the electrical grid (such as reserve), and could  dynamically adapt their consumption through demand response programs and more price-sensitive behaviours \cite{good2017reviewDemandResponse}. 
However, the deployment of such large quantity of energy at the local level will require the development of new design paradigms, with a change from centralised to decentralized schemes \cite{parag2016electricity}, and the adoption of new pricing mechanisms, particularly at the local level \cite{yuan2016distribution}.
Moreover, the current electricity network should be reconsidered, as the traditional design assumes that electrical power is generated by large generators connected upstream at the transmission level, and transmitted to final consumers downstream at the distribution level \cite{biggar2014economics}. 
By contrast, the future energy system will experience both active consumers and small local generators willing to trade energy among themselves, and to provide services and sell electricity up to main grid \cite{iacopoCommunityAPEN2019}. 
Therefore, both novel market mechanisms, and new network planning strategies, must be developed, to allow local consumers and distributed generations to be actively engaged with the future energy system.

In the literature, a seminal contribution in network planning is due to Paul Joskow and Jean Tirole \cite{joskow2005merchant}. They presented an extensive analysis of merchant transmission investments, with a focus on real-world issues, for example the effect of \textit{lumpiness} in transmission expansions, i.e. electric lines can be expanded only by discrete amounts (\add{notice that this problem also affects distribution lines}). Another crucial problem in network planning is the full recovery of investment costs, also termed the \textit{revenue adequacy condition}. In particular, the congestion rent\footnote{The congestion rent is defined as the price difference between line endpoints multiplied by the power flow.} collected by a network operator is not necessarily sufficient to guarantee the recovery of the \textit{fixed} investment costs \cite{kirschen2004fundamentals}. To overcome this issue, network charges are introduced \cite{iacopo2020exante}. 
 
\add{At the distribution grid level, network investment and operating costs are recovered by distribution network operators through distribution tariffs. These tariffs are differentiated in capacity-based and volumetric \cite{lu2018esigningDistributionNetwork}. 
Capacity-based tariffs usually depend either on the contractual installed capacity, or on the maximum utilized capacity, i.e. the peak power measured during a specific period (e.g. monthly, yearly). Tariffs based on contractual capacity are also termed ex-ante tariffs, as these are fixed once users are connected to the grid, whereas capacity tariffs depending on peak power are termed ex-post tariffs, as the actual monetary amount collected by the operator is determined only at the end of the considered period. 
Therefore, ex-post capacity tariffs increase the level of revenue uncertainty for the distribution network operator, but are more cost-reflective, as they can capture actual power usage patterns \cite{CEER2017distTariffs}. On the other side, volumetric tariffs depend on the energy consumed. These can be applied separately to the injected and withdrawn electricity, or can be net-metered (i.e. the tariff is applied to the resulting net energy utilized). Finally, fixed fees, e.g. standing service charges (£/year), may also be applied to recover residual network costs. Both capacity and volumetric tariffs can be charged according to different schemes, such as by using (i) a flat rate, which does not vary regardless of the actual energy consumed or capacity reached, (ii) a quantity-dependent rate, such as the case of increasing block tariffs \cite{yun2020IBT}, where the amount charged increases as the energy consumed (or power utilized) rises, and (iii) on a time-of-use basis, where different tariff amounts are charged at different time periods, e.g. day/night, summer/winter. 
An extensive discussion on regulatory principles for tariff design at the distribution grid level can be found in \cite{ortega2008distribution}, and \cite{grimm2020optimal} compares different schemes, including time-of-use pricing, critical-peak pricing, real-time pricing, and fixed price tariffs. 
A comprehensive review of distribution network tariffs currently applied in European countries is reported in \cite{af2015study}, whereas a comparison of capacity and volumetric tariff schemes based on a game-theory approach is reported in \cite{Schittekatte2018networkTariffs}.}

\add{Nowadays, there is an increasing consensus among regulators and academics \cite{CEER2017distTariffs, eu2016DSO, perez-arriaga2017aRegulatory} to increase the adoption of capacity-based tariffs, and to better incorporate time and  location  components  into network tariffs in order to  provide enhanced economic signals. In particular, the departure from volumetric tariffs in favour of capacity-based tariffs is especially important in the case of net-metered volumetric tariffs \cite{ed2013netMeterinHeadache}, as these ones may over-incentive distributed generation (e.g. the deployment of PV panels). 
Indeed, self-generated energy reduces the amount of electricity imported from the main grid, which therefore decreases the network tariff paid by these users under a net-metered scheme.
However, the presence of distributed generation does not necessarily reduce the overall costs sustained by the network operator (e.g. reinforcement costs).
This forces network operators to recover their costs from the remaining network users (e.g. non-flexible households), which can therefore actually experience an increase in their network tariffs, leading to a problem of fairness in network cost allocation \cite{Eid2014netMeteringPV}.}
 
The future energy system, where flexible consumers and small generators can dynamically adapt their consumption and generation as prices change, requires the adoption of efficient and responsive pricing schemes, in particular at the distribution level \cite{biggar2014economics}. 
However, despite the clear evidence that marginal nodal pricing provides a means to guarantee allocative efficiency \cite{green2007nodal}, several countries still model even their wholesale electricity markets as a single-zone market (e.g. Great Britain, France, Germany, Poland, Hungary \cite{bemvs2016bidding}). Furthermore, to the best of the authors' knowledge, nodal prices at the distribution level are not yet implemented anywhere.
This leads to a series of inefficiencies at the local level, since prices do not react to congestion by rising in importing zones and lowering in exporting zones. As a consequence, a distorted incentive for higher consumption and lower production can be observed, which can translate into greater balancing and reinforcement costs to guarantee grid stability and reliability \cite{biggar2014economics}. 
Two of the main reasons against the deployment of nodal prices are \cite{sahni2012beyond}: 
\begin{itemize}
	\item \textit{Granularity}\\
	Nodal prices can vary from one location to another even within a relatively small local area. This can be perceived as unfair by some consumers, in particular at the distribution level \cite{biggar2016networkPricing}. Indeed, distribution nodal prices can differ even within the same neighbourhood, as they reflect the actual congestion and power losses in the local grid.
	\item \textit{Volatility}\\
	Nodal prices can fluctuate significantly from one time period to another. This problem is particularly relevant for small consumers, as they are (i) unprepared for forecasting volatile prices, or (ii) even predicting their own future consumption \cite{fotouhi2017dynamic}. Therefore, volatile prices can expose them to an unacceptable degree of uncertainty.
\end{itemize}
For these reasons, some consumers prefer to pay fixed prices \cite{schlereth2018consumers}. By contrast, users equipped with smart meters and flexible technologies can benefit from time periods with favourable energy prices, and can increase their revenues by exploiting their flexibility, e.g. by providing reserve services to the grid \cite{ruokamo2019towards}.
\add{Regulators may also incentivise flexible pricing schemes to provide better economic signals, as this can induce efficient energy usage, and efficient investments in both generation and network assets \cite{biggar2014economics}.}
This motivates the need for new mechanisms such that nodal prices for flexible users and fixed prices for traditional consumers can coexist effectively at the distribution level.
In order to account for both groups, two types of users will be considered in this paper. 
\begin{enumerate}
	\item The first type is termed \textit{non-flexible consumers}. They do not want to be exposed to volatile market prices, and prefer to pay a fixed price. These consumers have inelastic demand curves, and collectively represent a large share of the energy demand currently traded in electricity markets \cite{rossetto2019assessing}. In particular, it has been estimated that 44\% of households have perfectly inelastic demand curves, i.e. their demand of electricity is indifferent to any change in market prices. This segment of the population is characterized by households that own no major electric appliance other than a refrigerator, and therefore do not have a significant opportunity to adjust when they use electricity \cite{reiss2005household}.
	\item  The second type refers to \textit{flexible users}, who have an opportunity to benefit from volatile prices. This class represents flexible consumers equipped with devices and new technologies capable of providing demand-response \cite{albadi2008summary}, as well as owners of small local generators and distributed energy resources \cite{olivella2018optimization}. In the proposed market design, these users are exposed to distribution nodal prices, and demand/offer flexible quantities (i.e. their demand/supply curves are elastic). In particular, they can offer services, such as reserve provision, to the grid by exploiting their flexibility. 
\end{enumerate}

To effectively enable the deployment of flexible resources at the local level, the distribution system operator (DSO) has to consider the additional (bidirectional) power flows generated by flexible users, in particular when making long-term network investment decisions to properly reinforce the distribution grid. For this reason, distribution expansion problems must explicitly consider the presence of flexible users into the investment decision process.

The aim of the proposed framework is to address emerging economic problems in local distribution grids related to the coexistence and interactions between non-flexible consumers, flexible users, and a DSO who manages the local electricity network. In particular, the main objectives can be summarized as follows:
\begin{itemize}
	\item First, we show how flexible users can provide reserve to the main grid under a nodal pricing scheme, while coexisting with non-flexible consumers paying fixed prices for their energy demand. The importance of nodal pricing relies on the efficient allocation of scarce resources including energy, line capacity, and technical constraints \cite{kirschen2004fundamentals}. In particular, marginal nodal pricing allows both the cost of congestion and the effect of power losses in distribution lines to be incorporated into electricity prices \cite{biggar2014economics}. Then, a fixed price for non-flexible consumers based on the underlying nodal prices is determined.
	\item Second, the DSO will optimally manage the distribution network, considering  bidirectional power flows from flexible users, by building new lines subject to fixed and variable costs. The recovery of the overall investment costs is guaranteed by determining an optimal network tariff levied on grid users, while accounting for congestion rent and network expansion costs. The electrical grid is modelled by using a second-order cone relaxation, whose solution is exact for radial networks under mild conditions \cite{gan2012exactWithW}, and fully accounts for both network losses and reactive power at the distribution grid level.
\end{itemize}
The proposed framework is structured as a non-linear integer bilevel model. The upper level problem represents an optimal long-term distribution grid planning problem, which determines the lines to be built or expanded, and the appropriate network tariff to be collected. The lower level program represents a market clearing problem spanning through the investment time horizon. The bilevel model is then recast as a single, equivalent mixed-integer quadratically constrained program (MIQCP) by resorting to complementarity relations and integer algebra, which can be solved with off-the-shelf solvers. To summarize, the main novelties of the proposed framework are:
\begin{enumerate}
\item determination of the optimal network tariff collected by the network operator and levied on distribution grid users, while accounting for congestion rent and expansion costs, and subject to the exact modelling of the distribution power flows by resorting to a second-order cone program (SOCP) relaxation;
\item reformulation of the bilevel model with SOCP constraints as a single equivalent MIQCP problem by using complementarity relations and integer algebra;
\item provision of flexibility (in terms of both active and reactive power) as reserve capacity to the main grid, furnished by flexible users;
\item computation of the fixed price charged to non-flexible consumers based on the underlying distribution nodal prices paid by flexible users, while ensuring that monetary payments are in equilibrium.
\end{enumerate}

The remaining sections are structured as follows. Section \ref{sec:model_description} describes the proposed bilevel model in detail. Section \ref{sec:fixedPrice} shows how the fixed price paid by non-flexible consumers is defined and computed. Section \ref{sec:resolution method} outlines how the proposed model can be recast as an equivalent MIQCP by resorting to complementarity relations and integer algebra. Section \ref{sec:numerical_results} reports three test cases based on a 5-bus network to highlight the main properties of the proposed approach, and one test case based on a 33-bus distribution grid. Finally, Section \ref{sec:conclusion} outlines the main conclusions. Variables and parameters used in the following sections are fully described in the Nomenclature in \ref{sec:nomenclature}.

\section{Optimal network tariffs and grid investment planning}\label{sec:model_description}

\subsection{Distribution grids}\label{sec:subsec:distribution_intro}

\begin{figure}[h!]
	\centering
	\includegraphics[width=0.6\linewidth]{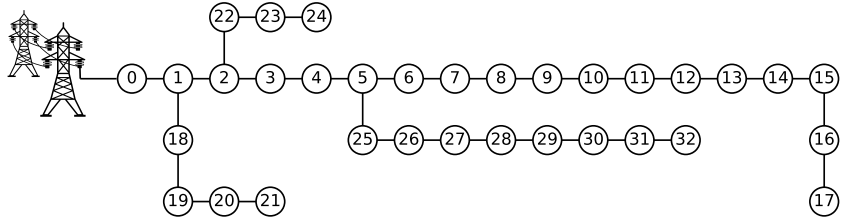}
	\caption{Distribution network with 33 buses (or nodes), where node zero represents the local power substation (which is connected upstream with the transmission grid), whereas the remaining nodes downstream depict the access points for local users connected to the distribution grid (e.g. households, flexible consumers, and generators).}
	\label{fig:33bus}
\end{figure}
Figure \ref{fig:33bus} shows an example of distribution grid. The local substation is located at node zero, and it is connected upstream with the transmission grid, whereas the remaining nodes downstream depict the access points for local users connected to the distribution grid (e.g. households, flexible consumers, and generators). In the proposed framework, flexible users can exploit their flexibility by selling upward and downward reserve capacity to the main grid. Upward reserve is the amount of additional power that can be injected in a node, whereas downward reserve is the further amount of power that can be withdrawn from a node. In the proposed model, upward reserve can be provided by generators using their spare power capacity, and by flexible consumers decreasing their consumption. Similarly, downward reserve can be provided by generators decreasing their power output, and by flexible consumers increasing their consumption. 

The approach required to model power flows in distribution grids differs significantly from the one usually employed in transmission networks. High-voltage transmission networks can be approximately described by using the so-called DC load flow method \cite{schavemaker2017electrical}. The key advantage of this approach is that power flows are represented by linear relations. It is based on three assumptions: (i) line resistances are negligible compared to reactances; (ii) the per-unit voltage magnitude at each node is one; (iii) and voltage phase angle differences across line endpoints are relatively small. These assumptions have important consequences. The first assumption implies that power losses can be neglected. The second one implies that reactive power flowing across lines is always zero. These assumptions approximately hold for high-voltage transmission lines, but do not hold in distribution grids. That is, both power losses and reactive power must be considered in describing low and medium voltage distribution networks. This requires non-linear modelling approaches. In the following, we consider distribution networks with a radial topology, as in Figure \ref{fig:33bus}. In this case, power flows at the distribution level can be represented by resorting to a SOCP relaxation, which ensures the exactness of the obtained solutions for these networks under mild technical assumptions \cite{gan2014exact, gan2012exactWithW}, which hold in all examples and test cases analysed in the following sections. 

\subsection{Bilevel programming}\label{sec:subsec:bilevel_intro}

The proposed framework leverages bilevel programming to determine the optimal grid investment planning and network tariffs at the distribution level. A bilevel model involves the optimization of two nested mathematical programs, termed upper and lower level problems \cite{bard1998practical}. Formally, a bilevel model is defined as follows:
\begin{subequations}\label{sketchBilevel}
	\begin{align}
	\max_{u \in \, \mathcal{U}(x^*)} &\,F(u,x^*)\label{sketchUpper}\\
	\text{s.t.}\,\,\,&x^* \in \arg \max_{x \in \mathcal{X}(u)} f(x;u) \, ,\label{sketchLower}
	\end{align}
\end{subequations}
where \eqref{sketchUpper} represents the upper level problem, whereas \eqref{sketchLower} characterizes the lower level problem. The functions $F$ and $f$ are their respective objective functions, and $\mathcal{U}$, $\mathcal{X}$ are constraint sets, which depend on the decision variables of the other level. A key property of bilevel programming is that all upper level decision variables, labelled as $u$ in \eqref{sketchUpper}-\eqref{sketchLower}, enter the lower level as fixed parameters. The variables $x^*$ represent the optimal solution of the lower level problem, which depends on the upper level variables $u$, i.e., $x^*\,{=x^*(u)}$. However, for ease of reading this dependence is usually not formally expressed. Historically, bilevel programming has been extensively used in the field of game theory to model Stackelberg games \cite{stackelberg1934}. However, in power system economics bilevel programs (as well as complementary problems \cite{conejo2012complementarity}), are not usually used to actually build a game, but as a means of accessing dual variables \cite{iacopoAPEN2018}, which are related to  electricity nodal prices under the marginal pricing framework \cite{schweppeCaramanis1982optimal, schweppe1988spot}. 

\begin{figure}[h!]
	\centering
	\includegraphics[width=0.5\textwidth]{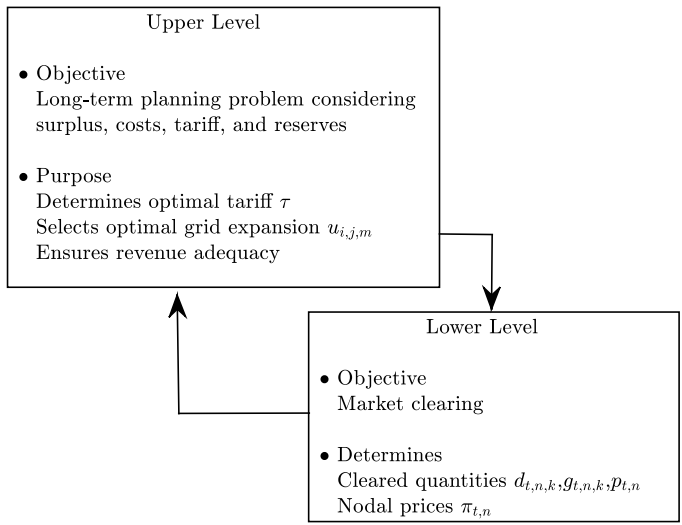}
	\caption{Bilevel framework. The upper level model represents a long-term investment planning problem accounting for participants' surplus, network tariff, investment costs, and revenues from reserve provision. It optimally determines the tariff $\tariff$ levied on network users, and the grid expansion level $\uBinary$. Given the selected grid expansion $\uBinary$, the lower level problem clears the market  to obtain the allocated demand and generation quantities $d_{t,n,k}$, $g_{t,n,k}$, the power injections $p_{t,n}$, and the market prices $\pi_{t,n}$, which are used in turn by the upper level to verify the revenue adequacy condition required to compute $\tariff$. The latter affects the selected grid expansion $\uBinary$ through the upper level objective function.}
	\label{fig:bilevel}
\end{figure}
Figure \ref{fig:bilevel} depicts the structure of the proposed bilevel framework. The upper level problem is a long-term investment planning problem, which accounts for participants' surplus, network tariff, reserve provision, and grid investment costs. It determines the optimal network tariff $\tariff$ collected by the network operator, and the optimal grid expansion $\uBinary$. In turn, the lower level solves a market clearing problem obtaining the optimal solution (labelled as $x^*$ in \eqref{sketchLower}), which includes the optimal allocated quantities $d_{t,n,k}$, $g_{t,n,k}$, the nodal power injections $p_{t,n}$, and the nodal prices $\pi_{t,n}$, for both active and reactive power. The bilevel formulation is required as the distribution network tariff $\tariff$ affects the upper level objective function, but it depends on the lower level nodal prices $\nodalPrice$, whereas the upper level influences the lower level through the selected network expansions $\uBinary$.

\subsection{Upper Level}\label{sec:subsec:upper_level}

This section describes the upper level problem introduced in Section \ref{sec:subsec:bilevel_intro}, which is defined as follows (see also the Nomenclature in \ref{sec:nomenclature}).
\begin{subequations}\label{upperLevelFull}
\begin{align}
\max_{ \tariff, \uBinary} \quad& \sum_{t \in \setTime}\sum_{n \in \setNodesPlus}\Big( \sum_{k \in \setDemNode} \priceDem {\dem}^* - \sum_{k \in \setGenNode} \priceGen {\gen}^* \Big)\Delta_t 
- \sum_{t \in \setTime} \Big(\priceActivePowerZero \activeInjZero^* + \priceReactivePowerZero \reactiveInjZero^* \Big)\Delta_t
\notag\\
&+ \sum_{t \in \setTime}\Big(\priceReserveUp \sum_{n \in \setNodesPlus} \sensitivityInjection {\reserveUp}^*
+ \priceReserveDown\sum_{n \in \setNodesPlus} \sensitivityInjection {\reserveDown}^*\Big)\Delta_t
\notag\\
&- \tariff \sum_{t \in \setTime}\sum_{n \in \setNodesPlus} \Big(\demFixed + \sum_{k \in \setDemNode}\demMax + \sum_{k \in \setGenNode} \genMax \Big)\Delta_t
\notag\\
&- \sum_{(i,j) \in \setLines}\sum_{m \in \setLumpy} \uBinary (\Kfix + \Kvar\Flumpy) \label{upperObj}\\
s.t.\notag\\
&- \sum_{t \in \setTime} \Big(\sum_{n \in \setNodes} {\nodalPrice}^*\activeInj^* 
+ \priceReactivePowerZero q_{t,0}^*\Big)\Delta_t
+ \tariff \sum_{t \in \setTime}\sum_{n \in \setNodesPlus} \Big(\demFixed + \sum_{k \in \setDemNode}\demMax + \sum_{k \in \setGenNode} \genMax \Big)\Delta_t \geq \notag\\
&\hspace{25em}\Kop + \sum_{(i,j) \in \setLines}\sum_{m \in \setLumpy} \uBinary (\Kfix + \Kvar\Flumpy) \label{upperRevAdequacy}
\end{align}
\begin{align}
&\sum_{(i,j) \in \setLines}\sum_{m \in \setLumpy} \uBinary (\Kfix + \Kvar\Flumpy) \leq \Ktot \label{budget_constraint}\\
&\sum_{m \in \setLumpy} \uBinary = 1 \label{upperBleq1}  \qquad\qquad \forall (i,j) \in \setLines\\
&\uBinary \in \mathcal{H} \label{upperUconstraintLinePath}\\
&\uBinary \in \{0,1\} \, , \quad \tariff \geq 0 \label{upperVarDomain}& \hspace{10em}&
\end{align}
\end{subequations}
The starred terms in \eqref{upperLevelFull} represent the optimal solution of the lower level problem, labelled as $x^*$ in \eqref{sketchBilevel}. The upper level objective function \eqref{upperObj} is composed by five groups of terms enclosed by parentheses, which represent in order: (i) the surplus of flexible consumers and producers, bidding different quantities at different prices $\priceDem$ and $\priceGen$, with $\Delta_t=1$ hour; (ii) the revenues (costs) from selling (buying) any excess (deficit) of active $\activeInjZero$ and reactive $\reactiveInjZero$ power to (from) the upstream transmission grid at the substation level, i.e. at node zero, at price $\priceActivePowerZero$ and $\priceReactivePowerZero$, respectively; (iii) the income from upward $\reserveUp$ and downward $\reserveDown$ reserve provision to the main grid, paid respectively $\priceReserveUp$ and $\priceReserveDown$; (iv) the total network tariff levied on all grid users (both non-flexible and flexible consumers, as well as generators) connected to the local distribution grid, paid proportionally to their access capacity (respectively $\demFixed$, $\demMax$, and $\genMax$); as well as (v) the fixed $\Kfix$ and variable $\Kvar$ investment costs to undertake the expansion $m$ on the line $(i,j)$ if a specific expansion is selected, i.e. if $\uBinary=1$. 
To summarize, the objective function maximizes the overall welfare including revenues from reserve provision, while minimizing the costs paid by the distribution operator and the costs sustained by network users. In particular, it ensures that the minimum tariff $\tariff$ is levied on grid users (see \cite{ortega2008distribution} for an extensive discussion on regulatory principles for distribution tariff design).
Discount factors are not included for ease of reading. As described in Section \ref{sec:introduction}, power lines can be augmented only by discrete amount, i.e. lumpy expansions, and the parameter $\Flumpy$ represents the additional discrete line capacity increase. As a consequence, the set $\setLumpy$ of possible capacity expansions is a finite set.
Constraint \eqref{upperRevAdequacy} enforces the \textit{revenue adequacy condition}. The first terms inside parentheses in the left-hand side represent the merchandising surplus \cite{kirschen2004fundamentals}, where the minus is due to the sign convention used to define the nodal injections as generation minus demand. The merchandising surplus represents the difference between the payments made by consumers and the revenues collected by generators, and is a source of revenues for the network operator. In lossless power flow models (e.g. DC load flow), the merchandising surplus coincides with the congestion rent \cite{biggar2014economics}. 
\add{The tariff $\tariff$ is a capacity-based tariff applied to maximum flexible demand $\demMax$ for flexible consumers, maximum output power $\genMax$ of sources for generators, and fixed demand $\demFixed$ for non-flexible consumers.}
Therefore, the condition \eqref{upperRevAdequacy} ensures that the income from merchandising surplus and collected tariffs is greater than or equal to the total investment costs paid by the network operator, where the term $\Kop$ accounts for any residual non-investment cost.  
Note that constraint \eqref{upperRevAdequacy} does not necessarily hold with an equality at the optimum, due to the presence of lumpy expansions (the interested reader is referred to \cite{iacopo2020exante} for a review of this aspect).
Constraint \eqref{budget_constraint} enforces an upper bound on the total financial expenses for grid investments, where $\Ktot$ is the total amount available for investments. Relation \eqref{upperBleq1} ensures that for each line $(i,j)$ only one expansion $m$ is undertaken, with $m \in \setLumpy = \{0, \ldots, 1\}$, where $m = 0$ represents the current line state (i.e. no expansion), whereas $m=1$ represents the expansion of $100$\% of the current capacity (i.e. the line is doubled). Finally, the relation \eqref{upperUconstraintLinePath} accounts for any additional constraints on investment decisions, dictated for example by environmental concerns, policies on investments, strategies on line expansion, or limits due to right-of-ways, where $\mathcal{H}$ is an abstract constraint set representing these conditions.\\

\subsection{Lower level}\label{sec:subsec:lowerLevel}

This section describes in detail the lower level problem sketched in Section \ref{sec:subsec:bilevel_intro}. The purpose of this optimization program is to solve a market clearing problem to obtain the allocated quantities $\dem$ and $\gen$, and the nodal prices $\nodalPrice$, while providing upward $\reserveUp$ and downward $\reserveDown$ reserve capacity to the main grid. The optimal power flows are modelled by resorting to a second-order cone relaxation, which is exact for radial networks under mild assumptions on both line impedances and voltage limits that hold in all reported test cases (see \cite{gan2012exactWithW, gan2014exact} for a technical discussion on  resistance, reactance, and voltage magnitude upper bound requirements). The total power injected in a single node, termed apparent power $s_{t,n} \in \mathbb{C}$ is a complex number, where the real part is termed active power $\activeInj \in \mathbb{R}$, whereas the imaginary part is termed reactive power $\reactiveInj \in \mathbb{R}$, that is, $s_{t,n} := \activeInj + \mathfrak{I} \reactiveInj$
where $\mathfrak{I}$ is the imaginary unit. Similarly, the voltage $V_{t,n}$ at each node $n$ is a complex number. To cast the optimization problem in the real domain, the following additional definitions are introduced:
\begin{align}
&\wii := V_{t,i} \bar{V}_{t,i} && \forallt, \forall i \in \setNodes\\
&W^{ij}_{t,i,j} := V_{t,i} \bar{V}_{t,j} = \wijP +\mathfrak{I} \wijQ && \forall t \in \setTime, \forall (i,j) \in \setLines\\
&W^{ji}_{t,j,i} := V_{t,j} \bar{V}_{t,i} = \wjiP +\mathfrak{I} \wjiQ && \forall t \in \setTime, \forall (i,j) \in \setLines\\
&\admitance := \aRealAdmittance -\mathfrak{I} \bImagAdmittance && \forall (i,j) \in \setLines, \forall m \in \setLumpy
\end{align} 
where $\admitance$ is the admittance of the line $(i,j)$ as a result of the expansion $m$, $V_{t,i}$ is the complex voltage in node $i$ at time $t$, the notation $\bar{V}_{t,i}$ refers to the complex voltage conjugate, $\wii \in \mathbb{R}$ is the voltage magnitude squared in node $i$, and $\wijP \in \mathbb{R}$, $\wijQ \in \mathbb{R}$, $\wjiP \in \mathbb{R}$, $\wjiQ \in \mathbb{R}$, $\aRealAdmittance \in \mathbb{R}$, and $\bImagAdmittance \in \mathbb{R}$, are auxiliary variables introduced to remove the complex terms. Note that the modelling approach for distribution grids, based on the second-order cone relaxation introduced in \cite{gan2012exactWithW, gan2014exact}, requires that the matrix:
\begin{equation}\label{semidefmatrix}
\begin{bmatrix} 
\wii & W^{ij}_{t,i,j} \\
W^{ji}_{t,j,i} &  \wjj
\end{bmatrix} 
\end{equation}
be positive semi-definite \cite{boyd2004convexOptimization}. By Sylvester criterion \cite{kocuk2016strong}, this condition is equivalent to requiring both $\wii \geq 0$ (which holds by definition), and:
\begin{align}\label{SOCPconstraint}
W^{ij}_{t,i,j} W^{ji}_{t,j,i} \leq \wii  \wjj &\iff (\wijP\big)^2 + \big(\wijQ\big)^2 + \Big(\dfrac{\wii - \wjj}{2}\Big)^2 \leq \Big(\dfrac{\wii + \wjj}{2}\Big)^2\notag\\
&\iff \left\lVert 
\begin{matrix}
\wijP\\
\wijQ\\
(\wii - \wjj)/2
\end{matrix}
\right\rVert_2 \leq \frac{\wii + \wjj}{2}
\end{align}
which is a second-order cone constraint, where $\frac{\wii + \wjj}{2} \geq 0$ as $W^{ii}_{t,(\cdot)} \geq 0$. Note that the matrix \eqref{semidefmatrix} is Hermitian by construction.\\

Considering the above definitions, and given the grid expansion $\uBinary$, the lower level problem introduced in Section \ref{sec:subsec:bilevel_intro} is defined as follows.
\begin{subequations}\label{lowerLevel}
\begin{flalign}
&({\dem}^*, {\gen}^*, \activeInjZero^*,  \reactiveInjZero^*, {\reserveUp}^*, {\reserveDown}^*, \activeInj^*,  [{\nodalPrice}^*]) = \notag\\
&\arg \max \, \sum_{t \in \setTime}\sum_{n \in \setNodesPlus}\Big( \sum_{k \in \setDemNode} \priceDem \dem - \sum_{k \in \setGenNode} \priceGen \gen \Big)
- \sum_{t \in \setTime} \priceActivePowerZero \activeInjZero 
- \sum_{t \in \setTime} \priceReactivePowerZero \reactiveInjZero\notag\\
&\qquad\qquad + \sum_{t \in \setTime} \Big(\priceReserveUp \sum_{n \in \setNodesPlus} \sensitivityInjection \reserveUp
+ \priceReserveDown\sum_{n \in \setNodesPlus} \sensitivityInjection \reserveDown\Big)& \label{lowerObjectiveFunction}
\end{flalign}
\begin{align}
&s.t.\notag\\
&\activeInj= \sum_{k \in \setGenNode} \gen - \sum_{k \in \setDemNode}\dem - \demFixed  && \forall t \in \setTime, \forall n \in \setNodesPlus&& [\nodalPrice \in \mathbb{R}] \label{lowerActiveInjDef}\\
&\reactiveInj= \sum_{k \in \setGenNode} \genReactive - \sum_{k \in \setDemNode} \demReactive  && \forall t \in \setTime, \forall n \in \setNodesPlus&& [\nodalPriceRective \in \mathbb{R}]  \label{lowerReactiveInjDef}
\end{align}
\begin{align}
&\activeInj = \sum_{(i,j) \in \setLines: i=n} \sum_{m \in \setLumpy} \uBinary \big( \wii \aRealAdmittance - \wijP \aRealAdmittance + \wijQ \bImagAdmittance \big)
+ \notag\\
&\qquad\sum_{(i,j) \in \setLines: j=n} \sum_{m \in \setLumpy} \uBinary \big( \wjj \aRealAdmittance - \wjiP \aRealAdmittance + \wjiQ \bImagAdmittance \big) && \forall t \in \setTime, \forall n \in \setNodes && [\lambdaPowerActive \in \mathbb{R}]  \label{lowerNodeAllActive}\\
&\reactiveInj = \sum_{(i,j) \in \setLines: i=n} \sum_{m \in \setLumpy} \uBinary \big( \wii \bImagAdmittance - \wijP \bImagAdmittance - \wijQ \aRealAdmittance \big) 
+ \notag\\
&\qquad\quad\sum_{(i,j) \in \setLines: j=n} \sum_{m \in \setLumpy} \uBinary \big( \wjj \bImagAdmittance - \wjiP \bImagAdmittance - \wjiQ \aRealAdmittance \big) && \forall t \in \setTime, \forall n \in \setNodes && [\lambdaPowerReactive \in \mathbb{R}]  \label{lowerNodeAllReactive}
\end{align}
\begin{align}
&\sum_{m \in \setLumpy} \uBinary \big(\wii \aRealAdmittance - \wijP \aRealAdmittance + \wijQ \bImagAdmittance \big)\leq \flowMax + \sum_{m \in \setLumpy} \uBinary \Flumpy &&
\forall t \in \setTime, \forall (i,j) \in \setLines && [\muMax \geq 0]  \label{lowerMaxFlow}\\
&\sum_{m \in \setLumpy} \uBinary \big(\wjj \aRealAdmittance - \wjiP \aRealAdmittance + \wjiQ \bImagAdmittance \big) \leq \flowMin + \sum_{m \in \setLumpy} \uBinary \Flumpy &&
\forall t \in \setTime, \forall (i,j) \in \setLines && [\muMin \geq 0]  \label{lowerMinFlow}
\end{align}
\vspace{-2em}
\begin{align}
&W^{ii}_{t,0} + 2 \sum_{(i,j) \in \mathcal{P}_n} \sum_{k: i \in \mathcal{P}_k} \sum_{m \in \setLumpy} \uBinary \big(\dfrac{\aRealAdmittance p_{t,k}}{\aRealAdmittance^2 + \bImagAdmittance^2}  + \dfrac{\bImagAdmittance q_{t,k}}{\aRealAdmittance^2 + \bImagAdmittance^2} \big) \leq (v^{max}_{t,n})^2 && \forallt, \forallnplus && [\beta_{t,n} \geq 0]\label{lowerW_hat} 
\end{align}
\vspace{-2em}
\begin{align}
&(\voltageMagnitudeMin)^2 \leq \wii \leq (\voltageMagnitudeMax)^2 && \forall t \in \setTime, \forall i \in \setNodesPlus && [\chiMin \geq 0, \chiMax \geq 0] \label{lowervoltagemax}\\
&\demMin\leq\dem \leq \demMax  && \forall t \in \setTime, \forall n \in \setNodesPlus, \forall k \in \setDemNode && [\phiActiveDemMin \geq 0,\phiActiveDemMax \geq 0] \label{lowerDemActiveMax}\\
&\genMin\leq \gen \leq \genMax  && \forall t \in \setTime, \forall n \in \setNodesPlus, \forall k \in \setGenNode && [\phiActiveGenMin \geq 0, \phiActiveGenMax \geq 0] \label{lowerGenActiveMax}\\
&\demReactiveMin\leq \demReactive \leq \demReactiveMax  && \forall t \in \setTime, \forall n \in \setNodesPlus, \forall k \in \setDemNode && [\phiReactiveDemMin \geq 0,\phiReactiveDemMax \geq 0] \label{lowerDemReactiveMax}\\
&\genReactiveMin\leq\genReactive \leq \genReactiveMax  && \forall t \in \setTime, \forall n \in \setNodesPlus, \forall k \in \setGenNode && [\phiReactiveGenMin \geq 0,\phiReactiveGenMax \geq 0] \label{lowerGenReactiveMax} \\
&\wijP = \wjiP && \forall t \in \setTime, \forall (i,j) \in \setLines && [\epsilonP \in \mathbb{R}] \label{lowerHermitanOne}\\
&\wijQ = - \wjiQ && \forall t \in \setTime, \forall (i,j) \in \setLines && [\epsilonQ \in \mathbb{R}]\label{lowerHermitanTwo}\\
&\reserveUp = \sum_{k \in \setGenNode} ( \genMax - \gen ) + \sum_{k \in \setDemNode} \dem  && \forall t \in \setTime, \forall n \in \setNodesPlus&& [\rhoUp \in \mathbb{R}]\label{lowerReserveUp}\\
&\reserveDown = \sum_{k \in \setGenNode} \gen + \sum_{k \in \setDemNode} ( \demMax - \dem)  && \forall t \in \setTime, \forall n \in \setNodesPlus&& [\rhoDown \in \mathbb{R}]\label{lowerReserveDown}
\end{align}
\vspace{-1.5em}
\begin{align}
&\big(\wijP\big)^2 + \big(\wijQ\big)^2 + \Big(\dfrac{\wii - \wjj}{2}\Big)^2 \leq \Big(\dfrac{\wii + \wjj}{2}\Big)^2 && 
\forall t \in \setTime, \forall (i,j) \in \setLines &&   \hspace{10em} \label{lowerSOCP}
\end{align}
\end{subequations}
The terms enclosed in squared brackets are dual variables. The lower level decision variables are $\dem \in \mathbb{R}$, $\gen \in \mathbb{R}$, $\demReactive \in \mathbb{R}$, $\genReactive \in \mathbb{R}$, $\activeInj \in \mathbb{R}$, $\reactiveInj \in \mathbb{R}$, $\reserveUp \in \mathbb{R}$, $\reserveDown\in \mathbb{R}$, $\wii \in \mathbb{R}$, $\wijP \in \mathbb{R}$, $\wijQ \in \mathbb{R}$, $\wjiP \in \mathbb{R}$, and $\wjiQ \in \mathbb{R}$. The objective function \eqref{lowerObjectiveFunction} maximizes the consumers and producers surplus, while accounting for the revenues from upward $\reserveUp$ and downward $\reserveDown$ reserve provision, and considering trades with the transmission grid at substation level. The positive (negative) active $\activeInjZero$ and reactive $\reactiveInjZero$ power represent the flow withdrawn from (injected into) the upstream transmission grid at time $t$. Constraints \eqref{lowerActiveInjDef}-\eqref{lowerReactiveInjDef} define the active and reactive power injections for each node $n \in \setNodesPlus$. Constraints \eqref{lowerNodeAllActive}-\eqref{lowerNodeAllReactive} enforce the Kirchhoff's first law, requiring the power injection to be equal to the sum of power inflows and outflows. Constraints \eqref{lowerMaxFlow}-\eqref{lowerMinFlow} set the limits on active power flows over each line $(i,j) \in \setLines$, where the amount $\Flumpy$ represents the additional capacity introduced by the expansion $\uBinary$. Note that the expansion affects also the line admittance through the terms $\aRealAdmittance$ and $\bImagAdmittance$. Constraints \eqref{lowerW_hat}-\eqref{lowervoltagemax} enforce voltage magnitude limits, where \eqref{lowerW_hat} is a technical condition to ensure that the obtained optimal power flows are exact (see Lemma 2 in \cite{gan2012exactWithW}). Constraints \eqref{lowerDemActiveMax}-\eqref{lowerGenReactiveMax} set demand and supply lower and upper bounds for flexible consumers and generators, for both active and reactive power. Constraints \eqref{lowerHermitanOne}-\eqref{lowerHermitanTwo} enforce the Hermitian property for the matrix \eqref{semidefmatrix}. Constraints \eqref{lowerReserveUp}-\eqref{lowerReserveDown} define both the upward and downward reserve, as described in Section \ref{sec:subsec:distribution_intro}. Finally, as shown in \eqref{SOCPconstraint}, the second-order cone constraint \eqref{lowerSOCP} enforces the positive semi-definite condition for matrix \eqref{semidefmatrix}. The (auxiliary) dual variables associated with the second-order cone components $\wijP$, $\wijQ$, $(\wii - \wjj)/2$, and $-(\wii + \wjj)/2$, are $\etaOne \in \mathbb{R}$, $\etaTwo \in \mathbb{R}$, $\etaThree \in \mathbb{R}$, and $\gammaSOCP \geq 0$, respectively. Note that, given $\uBinary$, the lower level problem is a second-order cone program, and therefore it is a convex optimization problem \cite{boyd2004convexOptimization}.

\begin{remark}
	The proposed bilevel framework is structured as a long-term planning model to determine the optimal network expansion, while (i) maximizing the overall social welfare net of investment costs, (ii) selecting the optimal network tariff (which ensures the revenue adequacy condition), and (iii) considering the collected merchandising surplus. 
	The DSO computes the latter by solving the lower level problem, which represents a centralized market clearing problem spanning the investment time horizon. 
	However, there is no requirement for the DSO to be the actual market operator. Indeed, the local market operator can be an independent entity \cite{pollitt2012lessons}, a for-profit entity \cite{iacopoCommunityAPEN2019}, or the DSO itself. In addition, the proposed framework can be generalized to consider several local markets  within the boundaries of the same distribution network. In this case, each local market can be represented as a distinct lower level-like problem, and the resulting model would be a mathematical problem with equilibrium constraints \cite{conejo2012complementarity}.
\end{remark}

\section{Determination of the fixed price}\label{sec:fixedPrice}

The fixed price $\priceUniqueDemFixed \in \mathbb{R}$ paid by non-flexible consumers in a local area $\mathcal{A}$ of the distribution grid, delimited by nodes $n\in \mathcal{A}\subseteq\setNodesPlus$, is defined by using the following relation:
	\begin{align}
	&\priceUniqueDemFixed \sum_{t \in \setTime}\sum_{n \in \mathcal{A}} \demFixed = \sum_{t \in \setTime}\sum_{n \in \mathcal{A}} \nodalPrice \demFixed \label{upperUniquePrice}
	\end{align}
	Condition \eqref{upperUniquePrice} defines the fixed price $\priceUniqueDemFixed$ charged to non-flexible consumers located in nodes $n\in \mathcal{A}$, as the average of the actual distribution nodal prices $\nodalPrice$, weighted by the fixed demand quantities $\demFixed$. Equivalently, the fixed price $\priceUniqueDemFixed$ is the price which ensures that the same total monetary amount is collected as if the non-flexible consumers paid nodal prices instead of the fixed one. By defining $\priceUniqueDemFixed$ as an average price, constraint \eqref{upperUniquePrice} implicitly generates a \textit{subsidy} between non-flexible consumers, where those at nodes with a lower nodal price pay more to compensate those at higher-priced nodes. Note that differences between nodal prices are due to line losses and network congestion. This means that by paying the same price $\priceUniqueDemFixed$, the effect of line losses and network congestion on non-flexible members within the same local area $\mathcal{A}$ is shared between them, which can be regarded as an example of fairness in local energy systems, if it is assumed that non-flexible consumers should not be penalised according to their location in the network. \add{Notice that non-flexible consumers can freely choose to opt out of this pricing scheme, and decide to pay nodal prices.} To summarize, the main features of the fixed price $\priceUniqueDemFixed$ defined through \eqref{upperUniquePrice} are:
	\begin{itemize}
		\itemsep0em
		\item it is based on the actual underlying nodal prices, which are not altered. This means that the \textit{allocative efficiency} induced by distribution nodal prices (computed in the lower level) is preserved \cite{biggar2016networkPricing};
		\item it ensures the \textit{overall monetary balance} at the system level, as long as \eqref{upperUniquePrice} holds;
		\item it generates a subsidising effect between non-flexible consumers within the same zone $\mathcal{A}$, who implicitly share the costs of power losses and network congestion, introducing an element of \textit{fairness} in the proposed approach.
	\end{itemize}
	Condition \eqref{upperUniquePrice} is added to the final MIQCP (reported in \ref{sec:Appendix:finalMIQCP}), obtained by recasting the bilevel model introduced in Section \ref{sec:model_description} as a single equivalent program, as shown in the following section.
	
	\begin{remark}
		Relation \eqref{upperUniquePrice} can be generalized by considering also some consumers $k \in \Omega^\mathcal{A}_{t,n}$ with elastic demand $\dem$ paying a fixed price, by recasting \eqref{upperUniquePrice} as follows:
		\begin{align}
		&\priceUniqueDemFixed \sum_{t \in \setTime}\sum_{n \in \mathcal{A}} \Big(\sum_{k \in \Omega^\mathcal{A}_{t,n}} \dem + \demFixed\Big) = \sum_{t \in \setTime}\sum_{n \in \mathcal{A}} \nodalPrice \Big(\sum_{k \in \Omega^\mathcal{A}_{t,n}} \dem + \demFixed\Big) \label{upperUniquePriceNonlin} \, ,
		\end{align}
		and by replacing any occurrence of $\nodalPrice$  with $\priceUniqueDemFixed$ in \eqref{dualDem}-\eqref{dualSOCP} for all $k \in \Omega^\mathcal{A}_{t,n}$, while substituting \eqref{strongDuality} with the complementarity slackness conditions for SOCP. However, \eqref{upperUniquePriceNonlin} is a non-convex relation, which leads to a non-convex integer problem, with no guarantee of the global optimum being found. The development of an efficient resolution method to handle this problem will be the subject of future work.
	\end{remark}

\section{Resolution method}\label{sec:resolution method}

\subsection{Single level problem}\label{sec:subsec:singleLevel}

\begin{figure}[h!]
	\centering
	\includegraphics[width=0.5\linewidth]{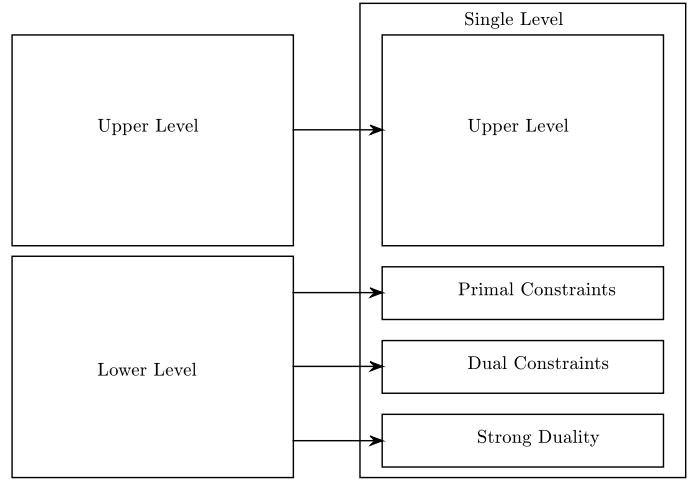}
	\caption{Reformulation scheme}
	\label{fig:single}
\end{figure}

When a lower level problem \eqref{sketchLower} is a convex optimization program, and satisfies at least one constraint qualification condition (such as the Slater's condition \cite{boyd2004convexOptimization}), it can be equivalently represented, within the bilevel model, by resorting to its first order necessary and sufficient Karush-Kuhn-Tucker (KKT) conditions. However, the KKT complementary slackness conditions are non-linear non-convex relations, which introduce non-linearities into the problem. To overcome this issue, a key optimization property can be exploited. The complementary slackness conditions hold if the strong duality property holds \cite{boyd2004convexOptimization}. Strong duality requires the equivalence between the objective function values of both the primal and dual problems. Therefore, as long as strong duality holds, a convex lower level problem \eqref{sketchLower} can be equivalently represented within the bilevel model \eqref{sketchBilevel} by resorting to its primal constraints, dual constraints, and the strong duality property, leading to a single and equivalent optimization problem \cite{conejo2012complementarity}, as outlined in Figure \ref{fig:single}.

The lower level problem introduced in Section \ref{sec:subsec:lowerLevel} is a second-order cone program, and therefore a convex optimization problem. In the following, we assume the Slater's condition holds, which is a reasonable assumption for any well-formed conic problem in real world applications (see \cite[p. 89]{mosek2019cookbook} for a discussion on this point). Therefore, for conic problems, the Slater's condition implies the strong duality property \cite{boyd2004convexOptimization}, which in turn implies the complementary slackness conditions \cite{lobo1998applications}. Therefore, the bilevel framework introduced in Section \ref{sec:subsec:bilevel_intro} can be recast as a single and equivalent optimization program, as follows:
\begin{subequations}\label{singleLevel}
\begin{align}
\max_{} \quad& \sum_{t \in \setTime}\sum_{n \in \setNodesPlus}\Big( \sum_{k \in \setDemNode} \priceDem \dem - \sum_{k \in \setGenNode} \priceGen \gen \Big) \Delta_t
- \sum_{t \in \setTime}\Big( \priceActivePowerZero \activeInjZero
+ \priceReactivePowerZero \reactiveInjZero\Big) \Delta_t
\notag\\
&
+ \sum_{t \in \setTime}\Big(\priceReserveUp \sum_{n \in \setNodesPlus} \sensitivityInjection \reserveUp
+ \priceReserveDown\sum_{n \in \setNodesPlus} \sensitivityInjection \reserveDown\Big)\Delta_t - \tariff \sum_{t \in \setTime}\sum_{n \in \setNodesPlus} \Big(\demFixed + \sum_{k \in \setDemNode}\demMax + \sum_{k \in \setGenNode} \genMax \Big)\Delta_t
\notag\\
&- \sum_{(i,j) \in \setLines}\sum_{m \in \setLumpy} \uBinary (\Kfix + \Kvar\Flumpy)\label{singleOBJ}\\
s.t.\notag\\
&\eqref{upperRevAdequacy}-\eqref{upperVarDomain} \label{singleUpperLevelConstraits}\\
&\eqref{lowerActiveInjDef}-\eqref{lowerSOCP} \label{singleLowerLevelPrimalConstraits}\\
&\eqref{dualDem}-\eqref{dualSOCP}\label{singleLowerLevelDualConstraits}\\
&\eqref{strongDuality} \label{singleStrongDuality}
\end{align}
\end{subequations}
where the objective function \eqref{singleOBJ} is the same as in the upper level problem \eqref{upperLevelFull}. Constraint \eqref{singleUpperLevelConstraits} represents the upper level constraints,  \eqref{singleLowerLevelPrimalConstraits} the lower level constraints, and \eqref{singleLowerLevelDualConstraits} its dual constraints (see \ref{sec:Appendix:dualConstraint}). Finally, \eqref{singleStrongDuality} refers to the strong duality property, reported in \ref{sec:Appendix:strongDuality}. Note that in the single level problem there is no distinction between upper and lower level variables. Therefore, the decision variables of the single level program \eqref{singleLevel} are the decision variables of the upper level, as well as those of the primal and dual lower level problems.

\subsection{Reformulation as MIQCP} \label{sec:refomulationMIQCP}

The single level problem \eqref{singleLevel} is a non-linear integer optimization problem.  The main sources of non-linearities are due to the presence of bilinear terms. In particular, there are two types of bilinear terms:
\begin{enumerate}
\item the product of the binary variable $\uBinary$ and a continuous variable, such as $\uBinary\wii$ in \eqref{lowerMaxFlow};
\item the product $\nodalPrice\activeInj$ in \eqref{upperRevAdequacy}.
\end{enumerate}
Note that the second-order cone constraints \eqref{lowerSOCP} and \eqref{dualSOCP} are convex relations, so they are not an issue. The first type of non-linearity can be removed by using standard integer algebra. Indeed, the product $u x$ involving the binary variable $u$ and the continuous bounded variable $-M \leq x \leq M$ can be equivalently replaced by introducing the auxiliary variable $y^{u x}$ defined as follows:
\begin{subequations}
\begin{align}
&-Mu \leq y^{ux} \leq Mu \label{aux1}\\
&-M(1-u) \leq x - y^{ux} \leq M(1-u) \label{aux2}
\end{align}
\end{subequations}
By using \eqref{aux1}-\eqref{aux2} all bilinear terms in the single level program \eqref{singleLevel} involving binary variables can be removed (a detailed description of the introduced auxiliary variables is reported in \ref{sec:Appendix:auxiliaryVariables}). In order to remove the non-linearity of the second type, some results from \cite{iacopo2020exante} (obtained by using the linear DC load flow approximation) have been extended to the second-order cone relaxation used in \eqref{lowerLevel}, resulting in the following equivalence, which holds at the optimum of the lower level problem (a proof is given in \ref{sec:Appendix:proof}):
	\begin{align}\label{lemma1}
	\nodalPrice\activeInj =& \sum_{k \in \setGenNode} (\phiActiveGenMax\genMax - \phiActiveGenMin\genMin + \priceReserveUp\sensitivityInjection\gen - \priceReserveDown\sensitivityInjection\gen + \priceGen\gen) \notag\\
	&- \sum_{k \in \setDemNode} (-\phiActiveDemMax\demMax + \phiActiveDemMin\demMin + \priceReserveUp\sensitivityInjection\dem - \priceReserveDown\sensitivityInjection\dem + \priceDem\dem) \notag\\
	&- \nodalPrice\demFixed \qquad\qquad \forall n \in \setNodesPlus
	\end{align}
and $\pi_{t,0}\activeInjZero = \priceActivePowerZero\activeInjZero$ by definition.\\

By using the equivalence \eqref{lemma1} and the auxiliary variables \eqref{auxiliaryStart}-\eqref{auxiliaryBetau}, the single level problem \eqref{singleLevel} can be equivalently recast as a mixed-integer quadratically constrained program (MIQCP), fully reported in \ref{sec:Appendix:finalMIQCP}, which can be solved with off-the-shelf solvers. \add{In addition, \ref{sec:stoch} outlines how the proposed bilevel model can be recast as a stochastic optimization problem to account for uncertainty e.g. in realizations of loads, prices, and renewable energy resources.}

\section{Numerical results and discussion}\label{sec:numerical_results}

\subsection{Settings}

This section describes test cases that are solved using the proposed MIQCP. The expansion $\uBinary = 1$ enforces the line expansion $m \in \setLumpy$ on the line $(i,j) \in \setLines$, through the following parameters:
\begin{align}
&\aRealAdmittance = a_{i,j,0}(1 + m) \label{aCost}\\
&\bImagAdmittance = e_{i,j,0}(1 + m) \label{bCost}\\
&\Flumpy = m \, \flowMax \label{flowCost}\\ 
&\Kfix = m \, K^{fix}_{i,j,1} \label{fixedCosts}\\
&\Kvar \Flumpy = m \, K^{var}_{i,j,1} \label{varCosts}\flowMax
\end{align}
where $a_{i,j,0}$ is the line conductance and $e_{i,j,0}$ the negative of the line susceptance, i.e., $y_{i,j,0} = a_{i,j,0} - \mathfrak{I} e_{i,j,0}$, where $y_{i,j,0}$ is the admittance before any expansion. The term $\flowMax$ is the maximum power flow over the line $(i,j)$ without expansion. The parameter $K^{fix}_{i,j,1}$ and $K^{var}_{i,j,1}$ represent the fixed and variable costs of building a  new line. Voltages are expressed in per-unit, with the substation voltage magnitude (at the slack bus)  equal to one, and maximum and minimum voltage magnitude limits of the remaining nodes set as $\voltageMagnitudeMax=1.20$ and $\voltageMagnitudeMin=0.80$, respectively. For ease of exposition, we assume a single time period, and one flexible consumer and one generator per node, i.e., $\setTime = \{1\}$, $\setDemNode = \{1\}$, and $\setGenNode = \{1\}$.
\add{Notice that, since the formulation of the lower level is convex, the computational burden will scale in polynomial time with the number of the continuous decision variables and time periods.} 
The MIQCP model has been implemented with Pyomo 5.6 \cite{pyomoBook}, and solved with Cplex 12.9 \cite{cplex2009v12} on a 64-core CPU with 256~GB of RAM.

\subsection{5-bus network}

\begin{figure}[h!]
	\centering
	\includegraphics[width=0.30\textwidth]{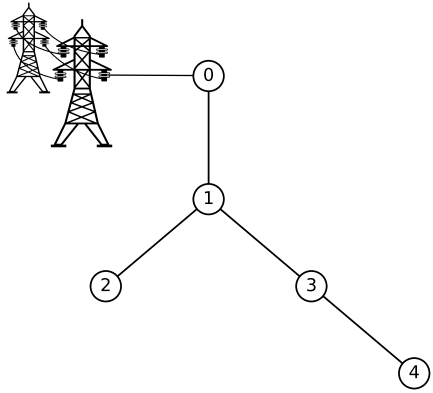}
	\caption{5-bus distribution grid.}
	\label{fig:5bus}
\end{figure}
This section reports three test cases based on the prototypical 5-bus grid depicted in Figure \ref{fig:5bus}. As described in Section \ref{sec:subsec:distribution_intro}, the substation is located at node zero (slack bus) and is connected upstream with the main transmission network.  In these examples, for each node $n \in \setNodesPlus$ with $\setNodesPlus = {1,2,3,4}$, the amount of fixed power required by non-flexible consumers (e.g. groups of households) is $\demFixed=100$~kW. Furthermore, flexible consumers demand up to $\demMax=100$~kW at each node. Similarly, generators offer up to $\genMax=100$~kW at each node.

\subsubsection{Reserve provision}\label{sec:subsec:numerical:5busReserve}

The first test case based on the 5-bus grid shows the effect of reserve provision on flexible consumers and generators. To focus on this effect, line expansions are not allowed, i.e. $\setLumpy = \{0\}$. Active power can be traded with the transmission network at price $\priceActivePowerZero = 5$~p/kWh, whereas the price of reactive power is set to zero, i.e., $\priceReactivePowerZero = 0$~p/kWh, and $\flowMax = 1,000$~kW for each line. 

\begin{table}[H]
	\centering
	\setlength{\extrarowheight}{2pt}
	\caption{5-bus network: effect of reserve provision on flexible users, with demand bid price $\priceDem$~p/kWh, generation bid price $\priceGen$~p/kWh, price of upward reserve $\priceReserveUp$~p/kWh, price of downward reserve $\priceReserveDown$~p/kWh, nodal prices $\nodalPrice$~p/kWh, allocated flexible demand $\dem$~kW, and allocated flexible generation $\gen$~kW.}
	\label{tab:caseReserveQuantity}
	\begin{tabular}{|c|c|c|l|r|r|l|r|r|}
		\hline
		\multirow{2}{*}{\begin{tabular}[c]{@{}c@{}}$n$\end{tabular}} & \multirow{2}{*}{\begin{tabular}[c]{@{}c@{}}$\priceDem$\end{tabular}} & \multirow{2}{*}{\begin{tabular}[c]{@{}c@{}}$\priceGen$\end{tabular}} & \multicolumn{3}{c|}{$\priceReserveUp=0$}                                                      & \multicolumn{3}{c|}{$\priceReserveUp=5$}                                                      \\ \cline{4-9} 
		&                                                                      &                                                                      & \multicolumn{1}{c|}{$\nodalPrice$} & \multicolumn{1}{c|}{$\dem$} & \multicolumn{1}{c|}{$\gen$} & \multicolumn{1}{c|}{$\nodalPrice$} & \multicolumn{1}{c|}{$\dem$} & \multicolumn{1}{c|}{$\gen$} \\ \hline
		1                                                                   & 40                                                                   & 20                                                                   & 30.36                     & 100                      & 100                      & 30.62                     & 100                      & 100                      \\ \hline
		2                                                                   & 35                                                                   & 25                                                                   & 30.54                     & 100                      & 100                      & 30.80                      & 100                      & 100                      \\ \hline
		3                                                                   & 30                                                                   & 30                                                                   & 30.51                     & 0                        & 100                      & 31.08                     & 100                      & 0                        \\ \hline
		4                                                                   & 25                                                                   & 35                                                                   & 30.67                     & 0                        & 0                        & 31.24                     & 0                        & 0                        \\ \hline
	\end{tabular}
\end{table}  
Table \ref{tab:caseReserveQuantity} shows the effect of increasing the price of upward reserve provision $\priceReserveUp$ from zero to $5$~p/kWh. The second and third columns report the demand bid price  $\priceDem$ (i.e. the marginal benefit, or willingness-to-pay) and the generation bid price $\priceGen$ (i.e. the marginal cost, assuming perfect competition)  at each node $n$, whereas the remaining columns show the obtained nodal prices $\nodalPrice$, and the cleared quantities $\dem$ and $\gen$ for flexible consumers and generators under the two different reserve prices. In both cases, no line is congested, and the differences in nodal prices are only due to the effect of power losses. For ease of reading, the definition of upward reserve in \eqref{lowerReserveUp} is reported again here:
\begin{equation}
\reserveUp = \sum_{k \in \setGenNode} ( \genMax - \gen ) + \sum_{k \in \setDemNode} \dem  
\end{equation}
That is, upward reserve can be provided by using the spare generator capacity  $\genMax - \gen$ to increase the power output, and by flexible consumers reducing their allocated consumption $\dem$. The important point to highlight in this example is the actual quantity $\dem$ and $\gen$ allocated in node $3$ (third row in Table \ref{tab:caseReserveQuantity}). 
\begin{itemize}
	\item Case $\priceReserveUp=0$~p/kWh\\
	When the upward reserve price is equal to zero, the nodal price in node $3$ is $30.51$~p/kWh. The demand bid price in that node is $30$~p/kWh. As a consequence, the demand bid order is out-of-the-money \cite{iacopoTPRWS2017}, and fully rejected (i.e. $\dem=0$). The generation bid price in node $3$ is also $30$~p/kWh. In this case, the supply bid order is in-the-money and fully executed (i.e. $\gen=100$), as it yields a profit of $0.51$~p/h for each kW.  
	\item Case $\priceReserveUp=5$~p/kWh\\
	When the price of upward reserve $\priceReserveUp$ rises from zero to $5$~p/kWh, the optimal allocation in node $3$ is the opposite of the previous case. That is, the supply order is fully rejected (i.e. $\gen=0$), and the demand order is fully accepted (i.e. $\dem=100$). In the previous case, the generator in node $3$ makes a profit of $0.51$~p/h for each kW traded. Now, by having its capacity unallocated, it can provide its full capacity as upward reserve, which is paid $5$~p/kWh, and therefore leads to a greater profit. Similarly, in the previous case the demand bid was fully rejected, whereas now the demand order is fully executed, even though its bid of $30$~p/kWh is smaller than the (new) nodal price of $31.08$~p/kWh. The reason is that flexible consumers also collect revenues from the upward reserve provision, which is paid $5$~p/kWh, leading to a surplus of $30-31.08+5=3.92$~p/h for each kW of potential consumption reduction. 
\end{itemize}
Moreover, as a consequence of the shift in allocated demand and generation, there is a net reduction of the power supplied, which leads to a nodal price increase, and induces a different allocation of surpluses between consumers and producers in the two cases. In detail, Table \ref{tab:caseReserveQuantity2} shows the surplus from power trades and the revenues from reserve provision, for both flexible consumers and generators. As can be observed, despite the loss in surplus due to the increase of nodal prices, the revenues from reserve provision allow flexible consumers and generators to achieve a significant welfare increase. However, the increase of nodal prices leads to an increase of the price $\priceUniqueDemFixed$ charged to non-flexible consumers, which is linked to nodal prices through \eqref{upperUniquePrice}. In this case, it rises from $30.52$~p/kWh to $30.93$~p/kWh, without being offset by reserve revenues (non-flexible consumers cannot provide reserve as their demand is fixed). 

This example shows how the possibility of providing reserve could represent a significant opportunity to make profits for flexible users in a local distribution network, and it should be regarded as an incentive for non-flexible consumers to adopt more price-sensitive behaviours and flexible technologies. 
\begin{table}[H]
	\centering
	\setlength{\extrarowheight}{2pt}
	\caption{5-bus network: effect of reserve provision on surplus and revenues (\pounds).}
	\label{tab:caseReserveQuantity2}
	\begin{tabular}{l|r|r|r|r|r|}
		\cline{2-6}
		\multicolumn{1}{c|}{}      & \multicolumn{1}{c|}{\begin{tabular}[c]{@{}c@{}}Surplus \\ flexible consumers\end{tabular}} & \multicolumn{1}{c|}{\begin{tabular}[c]{@{}c@{}}Revenues\\ reserve consumers\end{tabular}} & \multicolumn{1}{c|}{\begin{tabular}[c]{@{}c@{}}Surplus \\ flexible generators\end{tabular}} & \multicolumn{1}{c|}{\begin{tabular}[c]{@{}c@{}}Revenues \\ reserve generators\end{tabular}} & \multicolumn{1}{c|}{Total} \\ \hline
		\multicolumn{1}{|l|}{$\priceReserveUp=0$} & 14.1                                                                                          & 0                                                                                             & 16.4                                                                                           & 0                                                                                              & 30.5                       \\ \hline
		\multicolumn{1}{|l|}{$\priceReserveUp=5$} & 12.5                                                                                          & 15                                                                                         & 16.4                                                                                           & 10                                                                                           & 53.9                       \\ \hline
	\end{tabular}
\end{table}

\subsubsection{Network tariffs}\label{sec:subsec:numerical:5busInvestment}

The second test case based on the 5-bus grid shows the effect of investment decisions on network tariffs. In this example, the generation $\genMax$ is set to zero, whereas all other settings are same as the previous section, with maximum flow capacity $\flowMax=800$~kW. Reserve provision is not considered. The investment cost parameters are $K^{fix}_{i,j,1} = 100$~p, $K^{var}_{i,j,1} = 0.1$~p/kW, and $\Kop=0$. Two expansion sets $\setLumpy_1$ and $\setLumpy_2$ are compared. In the first set, the possible expansions are $\setLumpy_1 = \{0, 0.25, 0.50, 0.75, 1\}$, i.e., all network lines can be expanded with step increment equal to 25\% of their current capacity. In the second set, the possible expansions are $\setLumpy_2 = \{0, 0.50, 1\}$. 

\begin{table}[H]
	\centering
	\setlength{\extrarowheight}{2pt}
	\caption{Merchandising surplus, total network tariff, fixed and variable investment costs, and profit (p).}
	\label{tab:caseInvestment1}

\begin{tabular}{c|r|r|r|r|r|}
	\cline{2-6}
	& \multicolumn{1}{c|}{\begin{tabular}[c]{@{}c@{}}Merchandising \\ surplus\end{tabular}} & \multicolumn{1}{c|}{\begin{tabular}[c]{@{}c@{}}Total\\  tariff\end{tabular}} & \multicolumn{1}{c|}{\begin{tabular}[c]{@{}c@{}}Fixed \\ costs\end{tabular}} & \multicolumn{1}{c|}{\begin{tabular}[c]{@{}c@{}}Variable \\ costs\end{tabular}} & \multicolumn{1}{c|}{Profit} \\ \hline
	\multicolumn{1}{|c|}{$\setLumpy_1$} & 88                                                                                    & 0                                                                            & 25                                                                          & 20                                                                             & 43                          \\ \hline
	\multicolumn{1}{|c|}{$\setLumpy_2$} & 78                                                                                    & 12                                                                           & 50                                                                          & 40                                                                             & 0                           \\ \hline
\end{tabular}

\end{table}
The optimal solution shows that in both cases the line $(0,1)$ is expanded, with expansion $m=0.25$ in the first case, and $m=0.50$ in the second case. Table \ref{tab:caseInvestment1} shows the merchandising surplus and the total tariff collected by the network operator, as well as both fixed and variable investment costs, and the resulting profit. Fixed and variable costs are determined as in \eqref{flowCost}-\eqref{varCosts}. The total investment cost in the first case is equal to $45$~p, whereas the collected merchandising surplus is $88$~p, leading to a profit collected by the network operator equal to $43$~p. This extra profit is due to the lumpiness of network expansions, i.e. in real world instances, a line can only be increased by discrete amounts. This also shows that constraint \eqref{upperRevAdequacy} does not necessarily hold with an equality at the optimum (for an extensive discussion on this point see \cite{joskow2005merchant, iacopo2020exante}). In the second case, the expansion of $50$\% leads to a total investment cost of $90$~p. The merchandising surplus is only $78$~p, and therefore a tariff $\tariff$ equal to $0.014$~p/kW per hour is charged to network users, which allows the network operator to recover the remaining amount of $12$~p, with zero profit. 

\add{Notice that the tariff $\tariff$ is computed at the beginning of the investment period by using forecasts of energy mix and composition of customers. 
This may lead to a revenue shortfall if the actual values    significantly deviate from the ones used during the planning phase.
To overcome this issue, the network operator can recalibrate the tariff $\tau$ to ensure the recovery of the overall investment costs as follows. 
Suppose that after the investment in the line $(0,1)$ is made with lumpy capacity expansion $m=0.5$, the actual flexible users are half of those expected, leading to a shortfall of $50$ p.
To prevent this shortfall, the network operator can recalibrate the tariff by setting $K^{op}=50$ in (2b), and solve the proposed model again using the new information available. 
The resulting tariff will reflect the costs of both new and past investments.
In this example, no further lines are built, but a tariff of 0.011 p/kW per hour is still levied on network users to recover remaining investment costs.} 
This simple example shows how a distribution system operator can manage the local grid, by optimally selecting the lines to reinforce and the network tariff.

\subsubsection{\add{Different ratios of  flexible and non-flexible consumers}}\label{sec:subsec:numerical:5busRatio}

\add{This test case shows how different proportions of flexible and non-flexible consumers may affect allocated quantities, prices and social welfare. Both active and reactive power can be traded with the main transmission grid at $30$~p/kWh. The maximum power flow of the lines is $\flowMax=300$~kW, and no expansion is possible. 
Three scenarios are compared:
\begin{itemize}
	\item \textbf{Case 1}\\
	75\% non-flexible and 25\% flexible consumers (i.e. $\demFixed=150$ kW and $\demMax=50$ kW); 
	\item  \textbf{Case 2}\\
	50\% non-flexible and 50\% flexible consumers (i.e. $\demFixed=100$  kW and $\demMax=100$ kW); 
	\item \textbf{Case 3}\\
	25\% non-flexible and 75\% flexible consumers (i.e. $\demFixed=50$ kW and $\demMax=150$ kW). 
\end{itemize}}

\begin{table}[H]
	\centering
	\setlength{\extrarowheight}{2pt}
	\caption{5-bus network: effect of different proportions non-flexible and flexible consumers on prices (p/kWh)}
	\label{tab:caseDifferentProportionPrices}
	\begin{tabular}{c|c|c|c|}
		\cline{2-4}
		& \begin{tabular}[c]{@{}c@{}}Case 1 \\ (25\% flex.) \end{tabular} & \begin{tabular}[c]{@{}c@{}}Case 2 \\ (50\% flex.) \end{tabular} & \begin{tabular}[c]{@{}c@{}}Case 3 \\ (75\% flex.) \end{tabular} \\ \hline
		\multicolumn{1}{|c|}{$\pi^p_{t,1}$}          & 34.63                                                                    & 34.21                                                                      & 30.63                                                                     \\ \hline
		\multicolumn{1}{|c|}{$\pi^p_{t,2}$}          & 35.00                                                                    & 34.61                                                                      & 31.00                                                                     \\ \hline
		\multicolumn{1}{|c|}{$\pi^p_{t,3}$}          & 34.98                                                                    & 34.53                                                                      & 30.63                                                                     \\ \hline
		\multicolumn{1}{|c|}{$\pi^p_{t,4}$}          & 35.23                                                                    & 35.00                                                                      & 30.87                                                                     \\ \hline
		\multicolumn{1}{|c|}{$\priceUniqueDemFixed$} & 34.96                                                                    & 34.59                                                                      & 30.78                                                                     \\ \hline
	\end{tabular}
\end{table}  

\begin{table}[H]
	\centering
	\setlength{\extrarowheight}{2pt}
	\caption{5-bus network: effect of different proportions non-flexible and flexible consumers on allocated quantities (kW)}
	\label{tab:caseDifferentProportionQuantity}
	\begin{tabular}{|c|r|r|r|r|r|r|r|r|r|}
		\hline
		\multirow{2}{*}{\begin{tabular}[c]{@{}c@{}}Node\\ $n$\end{tabular}} & \multicolumn{3}{c|}{\begin{tabular}[c]{@{}c@{}}Case 1 \\(25\% flex.)\end{tabular}} & \multicolumn{3}{c|}{\begin{tabular}[c]{@{}c@{}}Case 2 \\(50\% flex.)\end{tabular}} & \multicolumn{3}{c|}{\begin{tabular}[c]{@{}c@{}}Case 3 \\(75\% flex.)\end{tabular}} \\ \cline{2-10} 
		& \multicolumn{1}{c|}{$\demFixed$}  & \multicolumn{1}{c|}{$\dem$} & \multicolumn{1}{c|}{$\gen$} & \multicolumn{1}{c|}{$\demFixed$}  & \multicolumn{1}{c|}{$\dem$}  & \multicolumn{1}{c|}{$\gen$} & \multicolumn{1}{c|}{$\demFixed$}  & \multicolumn{1}{c|}{$\dem$} & \multicolumn{1}{c|}{$\gen$} \\ \hline
		1                                                                   & 150                               & 50                          & 100                         & 100                               & 100                          & 100                         & 50                                & 150                         & 100                         \\ \hline
		2                                                                   & 150                               & 43                          & 100                         & 100                               & 100                          & 100                         & 50                                & 150                         & 100                         \\ \hline
		3                                                                   & 150                               & 0                           & 100                         & 100                               & 0                            & 100                         & 50                                & 0                           & 100                         \\ \hline
		4                                                                   & 150                               & 0                           & 100                         & 100                               & 0                            & 7                           & 50                                & 0                           & 0                           \\ \hline
		Tot.                                                                & 600                               & 93                          & 400                         & 400                               & 200                          & 307                         & 200                               & 300                         & 300                         \\ \hline
		
	\end{tabular}
\end{table} 
\add{For the three cases analysed, Table \ref{tab:caseDifferentProportionPrices} reports in the first four rows the nodal prices $\nodalPrice$ for each node, whereas the last row shows the fixed price $\priceUniqueDemFixed$ levied on non-flexible consumers. Table \ref{tab:caseDifferentProportionQuantity} shows the demanded quantities for non-flexible consumers $\demFixed$, the allocated demand $\dem$ of flexible consumers, and the cleared generation $\gen$ for each node (in the first four rows), and the total sum for the whole grid in the last row. As can be observed from Table \ref{tab:caseDifferentProportionPrices}, the increase of flexible consumption has a significant effect on prices. In particular, note how the increase of flexible demand leads to a reduction of the price $\priceUniqueDemFixed$ paid by the non-flexible consumers. This effect can be explained observing (see Table  \ref{tab:caseDifferentProportionQuantity}) that the flexible demand $\dem$ is never executed in node $3$ and $4$, where the nodal prices are significantly greater than the submitted bid $\priceDem$ which is $30$~p/kWh and $25$~p/kWh, respectively (see Table \ref{tab:caseReserveQuantity}). By contrast, non-flexible consumption must be always satisfied regardless of the price, which leads to a significant amount of power demanded in high-priced nodes. 
The line between node~$1$ and node~$0$ is congested in both Case 1 and Case 2, which contributes further to explain the greater nodal prices. 
The social welfare increase is equal to £$72.34$ from Case 1 (25\% flexible consumers) to Case 2 (50\% flexible consumers), and £$68.85$  from Case 2 to Case 3 (75\% flexible consumers).
This suggests that the presence of flexible consumers in a distribution network can have a positive impact in terms of increased welfare and lower prices.}

\subsection{33-bus network}\label{sec:subsec:num:33bus}

This test case is based on the IEEE 33-bus distribution network \cite{33bus1989networkIEEE}, which is shown in Figure \ref{fig:33bus}. Bid prices for flexible consumers and generators are sampled from normal distributions with standard deviations of $10$~p/kWh and means equal to $50$~p/kWh and $20$~p/kWh respectively. Negative prices are set to zero. Maximum flexible demand $\demMax$ and generation $\genMax$ are equal to half of the non-flexible demand $\demFixed$. Active and reactive power can be traded with the main grid at $30$~p/kWh. Reserve provision is not considered. The maximum power flow capacity of the lines is $\flowMax=2,000$~kW. The base voltage is $12.6$~kV and base apparent power is $1$~kVA. The investment cost parameters are $K^{fix}_{i,j,1} = 50$~\pounds, $K^{var}_{i,j,1} = 1$~p/kW, and $\Kop=0$. In this example, the constraint \eqref{upperUconstraintLinePath} is defined as:
\begin{align}\label{chain}
u_{i,j,1} \leq u_{h,k,1} && \forallij, \forall (h,k) \in \mathcal{P}_j \, ,
\end{align}
i.e. if a line $(i,j)$ is expanded, then all the upstream lines $(h,k)\in \mathcal{P}_j$ connected to this line up to the substation must be expanded. 
In this test case, two scenarios are compared. In the first one, no expansion is allowed. In the second one,  the set of the possible expansions is $\setLumpy = \{0, 1\}$, i.e. a line can be either doubled or left as it is. 

\begin{figure}[H]
	\centering
	\includegraphics[width=0.6\linewidth]{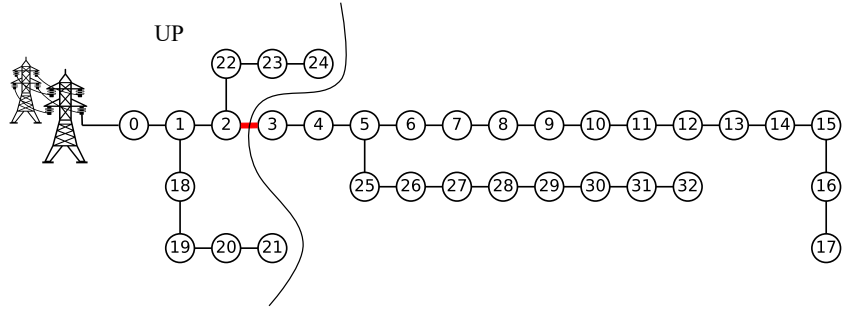}
	\caption{33-bus distribution grid. In the case with possible expansions $\setLumpy = \{0,1\}$, the line connecting nodes 2 and 3 is congested. The area labelled as ``UP'' includes all  nodes upstream of the congested line.}
	\label{fig:33busMarkedRed}
\end{figure}

\begin{table}[H]
	\centering
	\setlength{\extrarowheight}{2pt}
	\caption{Power flows over the first four lines in the 33-bus distribution network (kW)}
	\label{tab:case33bus}
	\begin{tabular}{|c|c|c|}
		\hline
		(i,j) & \setLumpy = \{0\} & \setLumpy = \{0, 1\} \\ \hline
		(0,1)  & 2,000              & 3,513                 \\ \hline
		(1,2)  & 1,715              & 3,047                 \\ \hline
		(2,3)  & 1,185              & 2,000                 \\ \hline
		(3,4)  & 1,119              & 1,868                 \\ \hline
		(…)   & …                 & …                    \\ \hline
	\end{tabular}
\end{table}
Table \ref{tab:case33bus} reports the power flows for the first four lines in the two scenarios. When no expansion is possible, the line $(0,1)$ is congested (power flow equal to $\flowMax=2,000$~kW). In the second case, both the line $(0,1)$ and $(1,2)$ are expanded, whereas the line $(2,3)$ is now the only one congested. 
\begin{figure}[H]
	\centering
	\includegraphics[width=0.9\linewidth]{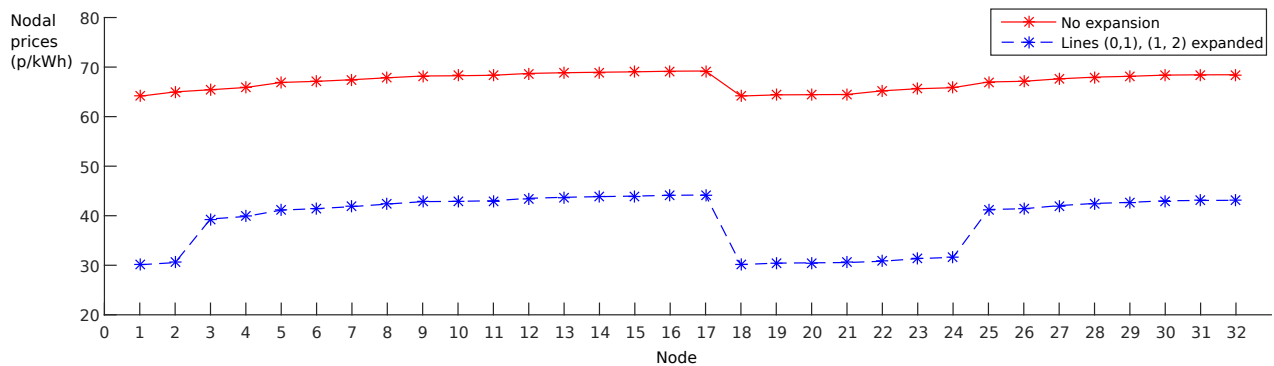}
	\caption{Nodal prices in the 33-bus distribution grid. The solid red line depicts the nodal prices in case of no expansion. The dashed blue line represents the nodal prices when the lines $(0,1)$ and $(1,2)$ are expanded.}
	\label{fig:33busPrices}
\end{figure}
Figure \ref{fig:33busPrices} shows the nodal prices in the two cases, where the solid red line depicts the no expansion case, whereas the dashed blue line represents the case when both the lines $(0,1)$ and $(1,2)$ are expanded. As can be observed, the nodal prices are significantly lower in the second case, in particular within the zone upstream of the congested line $(2,3)$, labelled as ``UP'' in Figure \ref{fig:33busMarkedRed}. 

\begin{figure}[H]
	\centering
	\includegraphics[width=0.6\linewidth]{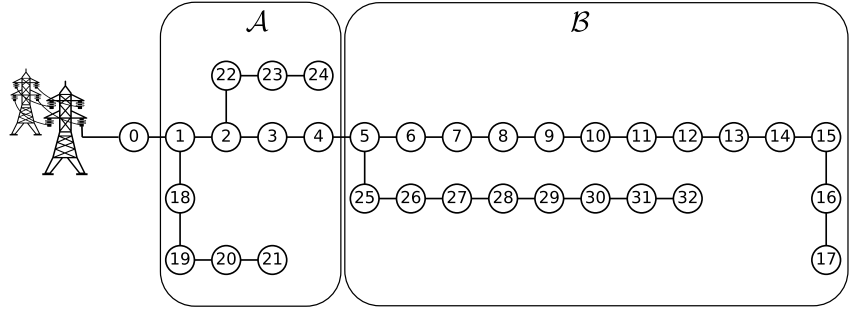}
	\caption{Areas $\mathcal{A}$ and $\mathcal{B}$. Within each area non-flexible consumers pay the same price, $\pi^D_\mathcal{A}$ and $\pi^D_\mathcal{B}$, respectively. The node zero represents the substation, and therefore it is not included in these areas.}
	\label{fig:33bus_AB}
\end{figure}

Now, consider two different areas in the grid, termed $\mathcal{A}$ and $\mathcal{B}$, and depicted in Figure \ref{fig:33bus_AB}. Suppose that non-flexible consumers in area  $\mathcal{A}$ pay the fixed price $\pi^D_\mathcal{A}$, whereas the ones in area  $\mathcal{B}$ pay the fixed price $\pi^D_\mathcal{B}$.
\noindent The network expansion leads to a decrease of the fixed price paid $\pi^D_\mathcal{A}$ from $65.26$~p/kWh to $31.94$~p/kWh in zone $\mathcal{A}$, and a reduction of the fixed price $\pi^D_\mathcal{B}$ from $68.09$~p/kWh to $42.68$~p/kWh in zone $\mathcal{B}$. In this case, the merchandising surplus generates sufficient revenues to recover the overall investment costs, and no tariff is applied. However, if further operation and management costs $\Kop=$£650 are considered, then a tariff equal to $0.42$~p/kW per hour have to be collected from all network users.

\begin{figure}[H]
	\centering
	\includegraphics[width=0.6\linewidth]{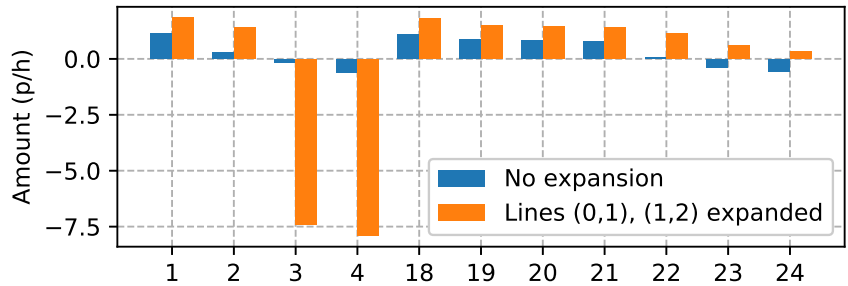}
	\caption{Subsidising effect between non-flexible consumers in zone $\mathcal{A}$, computed as the difference between the fixed price paid $\pi^D_\mathcal{A}$ and the actual nodal price in the consumer's location, i.e. $\pi^D_\mathcal{A} - \pi^p_{n}$ $\forall n \in \mathcal{A}$.}
	\label{fig:subsidy}
\end{figure}
Finally, Figure \ref{fig:subsidy} reports an example of subsidising effect between non-flexible consumers, by showing the difference between the fixed price $\pi^D_\mathcal{A}$ paid and the actual nodal price at each node in zone $\mathcal{A}$, i.e. $\pi^D_\mathcal{A}- \pi^p_{n}$ $\forall n \in \mathcal{A}$, for the cases with and without grid expansion. When the lines $(0,1)$ and $(1,2)$ are expanded, the only remaining congestion is between the nodes $2$ and $3$ (marked as red thick line in Figure \ref{fig:33busMarkedRed}). Due to this congestion, the nodal prices in nodes $3$ and $4$ rise to $39.34$~p/kWh and $39.85$~p/kWh, respectively, whereas the nodal prices in the remaining nodes in area $\mathcal{A}$ are at the most equal to $30.60$~p/kWh (see Figure \ref{fig:33busPrices}). However, in this case all non-flexible consumers pay the same fixed price $\pi^D_\mathcal{A}=31.94$~p/kWh. This means that the effect of the local congestion on prices is shared among all non-flexible consumers within the same area. This introduces a principle of fairness between non-flexible consumers in distribution grids, while allowing all remaining flexible users to use nodal prices.

\section{Conclusion}\label{sec:conclusion}

The increasing share of flexible users, who want to provide flexibility services to the grid, as well as the significant availability of renewable energy resources and distributed generation, motivate the introduction of nodal prices at the distribution level. Marginal nodal pricing is a desirable scheme as it leads to an efficient allocation of scarce resources in terms of both energy and line capacities. However, some consumers are unwilling to face nodal prices for two main reasons. First, nodal prices can fluctuate significantly from one time period to another (i.e. the volatility problem), which can expose non-flexible consumers to an unacceptable degree of uncertainty. Second, nodal prices can vary from one location to another even within the same neighbourhood (i.e. the granularity problem), which can be perceived as unfair. In this context, the proposed framework aims to shed light on how a distribution network could be optimally managed, by providing also new sources of profit for its users. In particular, it shows (i) how flexible and non-flexible consumers can coexist, where the former pay nodal prices and the latter are charged a fixed price based on the underlying nodal prices, (ii) how flexibility can be exploited by consumers and producers to obtain new streams of revenues by providing reserve capacity to the main grid, which could also incentivise non-flexible consumers to adopt more flexible technologies, (iii) how the distribution grid can be optimally expanded, while accounting for the additional bidirectional power flows from flexible users, and (iv) how distribution network tariffs can be optimally selected to ensure the recovery of the overall investment costs, while accounting for any merchandising surplus collected. Moreover, regulators could adopt the proposed mechanism to allow locational marginal pricing to be introduced incrementally at the distribution grid level, where customers could optionally sign up to use flexible prices, which would provide an incentive for the deployment of distribute energy resources at congested locations within the local network.

Future work will aim at (i) providing extensive policy analyses and recommendations based on the proposed framework, (ii) extending the presented model to consider upstream wholesale and ancillary service markets, (iii) introducing  consumers with elastic demand paying a fixed price, \add{and (iv) using a unit commitment problem as lower level, with uplifts to ensure the recovery of both start-up and variable costs}. In addition, further research lines include the reformulation of the proposed framework as a decentralized model (where the electrical grid is owned by different merchant investors that have to coordinate their investments). Moreover, the lower level market clearing problem can be extended to account for several local markets within the boundaries of the same distribution grid. \add{Finally, the proposed tariff could be differentiated according to several criteria, including location within the grid, the status of flexible/non-flexible users, or a reduced tariff could be charged to e.g. people experiencing energy poverty.}

\section*{Acknowledgement}

The present work has been supported by the EPSRC grant EP/S000887/1 and by the EPSRC grant EP/S031901/1, titled EnergyREV - Market Design for Scaling up Local Clean Energy Systems. Icons used are from Freepik by Flaticon.

\section*{Declaration of Interest}

The authors declare that there is no conflict of interest.

\appendix

\twocolumn

\section{Nomenclature}\label{sec:nomenclature}

\subsection*{\textbf{Sets and Indices}}
\begin{supertabular}{l p{0.82\columnwidth}}
	$\setLines$ & set of distribution lines\\
	$\setLumpy$ & set of possible lumpy capacity expansions for  distribution lines\\
	$\setNodes$ & set of distribution nodes, with $\setNodes = \{0, \ldots, N\}$, where $N$ is the total number of nodes\\
	$\setNodesPlus$ & set of all nodes except the slack bus (node zero), i.e. $\setNodesPlus=\setNodes \backslash \{0\}$\\
	$\setTime$ & set of time periods\\
	$\setDemNode$ & set of flexible consumers at time $t$ in node $n$\\
	$\setGenNode$ & set of generators at time $t$ in node $n$\\
\end{supertabular}

\subsection*{\textbf{Parameters}}
\begin{supertabular}{l p{0.82\columnwidth}}
	$\aRealAdmittance$ & conductance of the line $(i,j) \in \setLines$, in case of expansion $m \in \setLumpy$, p.u.\\
	$\priceActivePowerZero$ & price of active power traded with the transmission grid at time $t$, p/kWh\\
	$\priceReactivePowerZero$ &  price of reactive power traded with the transmission grid at time $t$, p/kVArh\\
	$\priceReserveUp$ & price of upward reserve at time $t$, p/kWh\\
	$\priceReserveDown$ & price of downward reserve at time $t$, p/kWh\\
	$\priceDem$ & demand bid price (willingness-to-pay) for flexible consumers $k$ in node $n$ at time $t$, p/kWh\\
	$\priceGen$ & supply bid price (marginal cost) for generators $k$ in node $n$ at time $t$, p/kWh\\
	$\demMax$ & maximum quantity of active power demanded by flexible consumer $k$ in node $n$ at time $t$, kW\\
	$\demMin$ & minimum quantity of active power demanded by flexible consumers $k$ in node $n$ at time $t$, kW\\
	$\demReactiveMax$ & maximum quantity of reactive power demanded by flexible consumers $k$ in node $n$ at time $t$, kVAr\\
	$\demReactiveMin$ & minimum quantity of reactive power demanded by flexible consumers $k$ in node $n$ at time $t$, kVAr\\
	$\demFixed$ & fixed demand required by non-flexible consumers in node $n$ at time $t$, kW\\
	$\bImagAdmittance$ & negative of the susceptance for the line $(i,j) \in \setLines$, in case of expansion $m \in \setLumpy$, p.u.\\
	$\Flumpy$ & lumpy capacity expansion $m \in \setLumpy$ for line $(i,j) \in \setLines$, kW\\
	$\flowMax$ & maximum power flow leaving node $i$ over the line $(i,j)$, kW\\
	$\flowMin$ & maximum power flow leaving node $j$ over the line $(i,j)$, kW\\
	$\genMax$ & maximum quantity of active power produced by generators $k$ in node $n$ at time $t$, kW\\
	$\genMin$ & minimum quantity of active power produced by generators $k$ in node $n$ at time $t$, kW\\
	$\genReactiveMax$ & maximum quantity of reactive power produced by generators $k$ in node $n$ at time $t$, kVAr\\
	$\genReactiveMin$ & minimum quantity of reactive power produced by generators $k$ in node $n$ at time $t$, kVAr\\
	$\Kvar$ & variable cost of expansion $m \in \setLumpy$ for the line $(i,j) \in \setLines$, p/kW\\
	$\Kfix$ & fixed cost of expansion $m \in \setLumpy$ for the line $(i,j) \in \setLines$, £ \\
	$\Kop$  & residual/non-investment costs paid by the distribution operator during the considered time periods, £\\
	$\Ktot$ & total financial budget for investments, £\\
	$\voltageMagnitudeMax $ & maximum voltage magnitude in node $n$ at time $t$, p.u.\\
	$\voltageMagnitudeMin $ & minimum voltage magnitude in node $n$ at time $t$, p.u.\\
	$\Delta_t$ & time span, equal to 1 hour.\\	
	%	$\sensitivityInjection$ & sensitivity parameter accounting for the location of node $n$ with respect to node zero\\
\end{supertabular}

\subsection*{\textbf{Variables}}
\begin{supertabular}{l p{0.82\columnwidth}}
	$\dem$ & allocated active power demand for flexible consumer $k$ in node $n$ at time $t$, kW\\
	$\demReactive$ & allocated reactive power demand  for flexible consumer $k$ in node $n$ at time $t$, kVAr\\
	$\gen$ & allocated active power for generator $k$ in node $n$ at time $t$, kW\\
	$\genReactive$ & allocated reactive power for generator  $k$ in node $n$ at time $t$, kVAr\\
	$\activeInj$ & active power injection in node $n$ at time $t$, kW\\
	$\reactiveInj$ & reactive power injection in node $n$ at time $t$, kVAr\\
	$\reserveUp$ & upward reserve provided in node $n$ at time $t$, kW\\
	$\reserveDown$ & downward reserve provided in node $n$ at time $t$, kW\\
	$\uBinary$ & binary variable equal to one if the expansion $m \in \setLumpy$ is performed on the line $(i,j) \in \setLines$\\
	$\wii$ & voltage magnitude squared at node $i$ at time $t$\\
	$\wijP$ & real component of the product $V_i \bar{V}_j$, where $V_i$ is the voltage at node $i$ and  $\bar{V}_j$ the complex voltage conjugate at node $j$\\
	$\wijQ$ & imaginary component of the product $V_i \bar{V}_j$, where $V_i$ is the voltage at node $i$ and  $\bar{V}_j$ the complex voltage conjugate at node $j$\\
	$\tariff$ & network tariff, p/kW per hour\\
	$\priceUniqueDemFixed$ & fixed price charged to non-flexible consumers, p/kWh\\
	$\nodalPrice$ & nodal price for active power, p/kWh\\
\end{supertabular}

\onecolumn

\section{Dual constraints}\label{sec:Appendix:dualConstraint}

This section reports the dual constraints of the lower level problem described in Section \ref{sec:subsec:lowerLevel}.
\begin{align}
&\nodalPrice + \phiActiveDemMax - \phiActiveDemMin - \rhoUp + \rhoDown = \priceDem && \forall t \in \setTime, \forall n \in \setNodesPlus, \forall k \in \setDemNode && [\dem \in \mathbb{R}] \label{dualDem}\\
&\nodalPriceRective + \phiReactiveDemMax - \phiReactiveDemMin = 0 && \forall t \in \setTime, \forall n \in \setNodesPlus, \forall k \in \setDemNode && [\demReactive \in \mathbb{R}]\\
&-\nodalPrice + \phiActiveGenMax - \phiActiveGenMin + \rhoUp - \rhoDown = - \priceGen && \forall t \in \setTime, \forall n \in \setNodesPlus, \forall k \in \setGenNode && [\gen \in \mathbb{R}] \label{dualGen}\\
&-\nodalPriceRective + \phiReactiveGenMax - \phiReactiveGenMin = 0 && \forall t \in \setTime, \forall n \in \setNodesPlus, \forall k \in \setGenNode && [\genReactive \in \mathbb{R}]
\end{align}
\begin{align}
&\lambda^p_{t,0} = - \priceActivePowerZero && \forall t \in \setTime && [\activeInjZero \in \mathbb{R}] \label{dualActiveInjZero} \\
&\lambda^q_{t,0} = - \priceReactivePowerZero && \forall t \in \setTime && [\reactiveInjZero \in \mathbb{R}] \label{dualReactiveInjZero}\\
&\nodalPrice + \lambdaPowerActive + 2 \sum_{(i,j) \in \mathcal{P}_n} \sum_{k: i \in \mathcal{P}_k} \sum_{m \in \setLumpy} \uBinary \dfrac{\aRealAdmittance}{\aRealAdmittance^2 + \bImagAdmittance^2} \beta_{t,k} = 0 && \forall t \in \setTime, \forallnplus && [\activeInj \in \mathbb{R}] \label{dualNodeActiveAll}\\
&\nodalPriceRective + \lambdaPowerReactive + 2 \sum_{(i,j) \in \mathcal{P}_n} \sum_{k: i \in \mathcal{P}_k} \sum_{m \in \setLumpy} \uBinary \dfrac{\bImagAdmittance}{\aRealAdmittance^2 + \bImagAdmittance^2} \beta_{t,k} = 0 && \forall t \in \setTime, \forallnplus && [\reactiveInj \in \mathbb{R}] \label{dualNodeReactiveAll}\\
&\rhoUp =  \priceReserveUp\sensitivityInjection && \forall t \in \setTime, \forall n \in \setNodesPlus && [\reserveUp \in \mathbb{R}]\label{dualReserveUp}\\
&\rhoDown =  \priceReserveDown\sensitivityInjection && \forall t \in \setTime, \forall n \in \setNodesPlus && [\reserveDown \in \mathbb{R}] \label{dualReserveDown}
\end{align}
\begin{align}
&\sum_{m \in \setLumpy} \uBinary \Big(\aRealAdmittance\lambdaPowerActiveI + \bImagAdmittance \lambdaPowerReactiveI - \aRealAdmittance \muMax\Big) + \epsilonP + \etaOne = 0  && \forallt, \forallij && [\wijP \in \mathbb{R}]\\
&\sum_{m \in \setLumpy} \uBinary \Big(\aRealAdmittance\lambdaPowerActiveJ + \bImagAdmittance \lambdaPowerReactiveJ - \aRealAdmittance \muMin\Big) - \epsilonP  = 0  && \forallt, \forallij && [\wjiP \in \mathbb{R}]\\
&\sum_{m \in \setLumpy} \uBinary \Big(-\bImagAdmittance \lambdaPowerActiveI + \aRealAdmittance \lambdaPowerReactiveI + \bImagAdmittance \muMax\Big) + \epsilonQ + \etaTwo  = 0  && \forallt, \forallij && [\wijQ \in \mathbb{R}]\\
&\sum_{m \in \setLumpy} \uBinary \Big(-\bImagAdmittance \lambdaPowerActiveJ + \aRealAdmittance \lambdaPowerReactiveJ + \bImagAdmittance
\muMin\Big) + \epsilonQ  = 0  && \forallt, \forallij && [\wjiQ \in \mathbb{R}]
\end{align}
\begin{align}
&\sum_{(i,j) \in \setLines: i=n} \Big( \sum_{m \in \setLumpy} \uBinary \big(- \aRealAdmittance \lambdaPowerActiveI - \bImagAdmittance \lambdaPowerReactiveI + \muMax \aRealAdmittance \big)
+ \dfrac{\etaThree - \gammaSOCP}{2}\Big) \notag\\ 
&+\sum_{(i,j) \in \setLines: j=n} \Big( \sum_{m \in \setLumpy} \uBinary\big(- \aRealAdmittance \lambdaPowerActiveJ - \bImagAdmittance \lambdaPowerReactiveJ + \muMin \aRealAdmittance\big) - \dfrac{\etaThree + \gammaSOCP}{2}\Big)\notag\\
& + \chiMax - \chiMin = 0 && \forallt,\forallnplus && [W^{ii}_{t,n} \in \mathbb{R}]\\
&(\etaOne)^2 + (\etaTwo)^2 + (\etaThree)^2 \leq (\gammaSOCP)^2 && \forallt, \forallij \label{dualSOCP}
\end{align}

\section{Strong Duality}\label{sec:Appendix:strongDuality}

The strong duality property requires the equivalence between the primal and dual objective function values, and it is defined as follows. 
\begin{flalign}
&\sum_{t \in \setTime}\sum_{n \in \setNodesPlus}\Big( \sum_{k \in \setDemNode} \priceDem \dem - \sum_{k \in \setGenNode} \priceGen \gen \Big)
- \sum_{t \in \setTime} \priceActivePowerZero \activeInjZero 
- \sum_{t \in \setTime} \priceReactivePowerZero \reactiveInjZero\notag\\
&\qquad\qquad + \sum_{t \in \setTime} \Big(\priceReserveUp \sum_{n \in \setNodesPlus} \sensitivityInjection \reserveUp
+ \priceReserveDown\sum_{n \in \setNodesPlus} \sensitivityInjection \reserveDown\Big) = \sum_{t \in \setTime}\sum_{n \in \setNodesPlus} - \nodalPrice\demFixed + \chiMax(\voltageMagnitudeMax)^2 - \chiMin(\voltageMagnitudeMin)^2 \notag\\
& + \sum_{t \in \setTime}\sum_{n \in \setNodesPlus}\big(\sum_{k \in \setDemNode}  \demMax\phiActiveDemMax + \demReactiveMax\phiReactiveDemMax - \demMin\phiActiveDemMin - \demReactiveMin\phiReactiveDemMin  +\rhoDown\demMax \big) \notag\\
& + \sum_{t \in \setTime}\sum_{n \in \setNodesPlus}\big(\sum_{k \in \setGenNode}  \genMax\phiActiveGenMax + \genReactiveMax\phiReactiveGenMax - \genMin\phiActiveGenMin - \genReactiveMin\phiReactiveGenMin + \rhoUp\genMax \big) \notag\\
&+ \sum_{t \in \setTime}\sum_{(i,j) \in \setLines} \flowMax\muMax + \flowMin\muMin + \sum_{m \in \setLumpy} \uBinary\Flumpy(\muMax + \muMin)  \notag\\
&+ \sum_{t \in \setTime}\sum_{(i,j) \in \setLines: i=0} \sum_{m \in \setLumpy} \wzerozero\uBinary \big( \aRealAdmittance \lambdaPowerActiveI + \bImagAdmittance \lambdaPowerReactiveI - \aRealAdmittance \muMax \big) \notag\\
&+ \sum_{t \in \setTime}\sum_{(i,j) \in \setLines: j=0} \sum_{m \in \setLumpy} \wzerozero\uBinary \big( \aRealAdmittance \lambdaPowerActiveJ + \bImagAdmittance \lambdaPowerReactiveJ - \aRealAdmittance \muMin \big) \notag\\
&+ \sum_{t \in \setTime}\sum_{n \in \setNodesPlus} \big((v^{max}_{t,n})^2 - W^{ii}_{t,0}\big)\beta_{t,n}\notag\\
&+ \sum_{t \in \setTime}\sum_{(i,j) \in \setLines: i=0} \wzerozero  \dfrac{- \etaThree + \gammaSOCP}{2} + \sum_{t \in \setTime}\sum_{(i,j) \in \setLines: j=0} \wzerozero  \dfrac{\etaThree + \gammaSOCP}{2}
\label{strongDuality}
\end{flalign}
where the voltage magnitude squared at the slack bus is $\wzerozero=1$~p.u., and bilinear terms, e.g. $\uBinary\muMax$, can be removed as detailed in Section \ref{sec:refomulationMIQCP}.

\section{Auxiliary variables}\label{sec:Appendix:auxiliaryVariables}

This section reports the auxiliary variables used to recast the single level problem \eqref{singleLevel} as a MIQCP program.

\begin{align}
&-(\voltageMagnitudeMax)^2 \uBinary \leq \yWijPu \leq (\voltageMagnitudeMax)^2 \uBinary \label{auxiliaryStart} \\
&-(\voltageMagnitudeMax)^2 (1-\uBinary) \leq \wijP - \yWijPu\leq (\voltageMagnitudeMax)^2 (1-\uBinary)\\
&-(\voltageMagnitudeMax)^2 \uBinary \leq \yWijQu \leq (\voltageMagnitudeMax)^2 \uBinary\\
&-(\voltageMagnitudeMax)^2 (1-\uBinary) \leq \wijQ - \yWijQu \leq (\voltageMagnitudeMax)^2 (1-\uBinary)\\
&-(\voltageMagnitudeMax)^2 \uBinary \leq \yWjiPu \leq (\voltageMagnitudeMax)^2 \uBinary\\
&-(\voltageMagnitudeMax)^2 (1-\uBinary) \leq \wjiP - \yWjiPu \leq (\voltageMagnitudeMax)^2 (1-\uBinary)\\
&-(\voltageMagnitudeMax)^2 \uBinary \leq \yWjiQu \leq (\voltageMagnitudeMax)^2 \uBinary\\
&-(\voltageMagnitudeMax)^2 (1-\uBinary) \leq \wjiQ - \yWjiQu \leq (\voltageMagnitudeMax)^2 (1-\uBinary)
\end{align}
\begin{align}
&0 \leq \ymuMaxU \leq M_1 \uBinary\\
&0 \leq \muMax - \ymuMaxU \leq M_1 (1-\uBinary)\\
&0 \leq \ymuMinU \leq M_1 \uBinary\\
&0 \leq \muMin - \ymuMinU \leq M_1 (1-\uBinary) \label{auxiliaryMuMi}\\
&(\voltageMagnitudeMin)^2 \uBinary \leq \yWiiu \leq (\voltageMagnitudeMax)^2 \uBinary\label{auxiliaryWii}\\
&(\voltageMagnitudeMin)^2 (1-\uBinary) \leq \wii  - \yWiiu \leq (\voltageMagnitudeMax)^2 (1-\uBinary) \\
&-M_1 \uBinary \leq \ylambdaPu \leq M_1 \uBinary\\
&-M_1 (1-\uBinary) \leq \lambdaPowerActiveI - \ylambdaPu\leq M_1 (1-\uBinary)\\
&-M_1 \uBinary \leq \ylambdaQu \leq M_1 \uBinary\\
&-M_1 (1-\uBinary) \leq \lambdaPowerReactiveI - \ylambdaQu\leq M_1 (1-\uBinary)\label{auxiliaryLambdaQu}\\
&M_2 \uBinary \leq \yPU \leq M_3 \uBinary\label{auxiliaryPu}\\
&M_2 (1-\uBinary) \leq p_{t,k} - \yPU \leq M_3 (1-\uBinary)\\
&M_4 \uBinary \leq \yQU \leq M_5 \uBinary\\
&M_4 (1-\uBinary) \leq q_{t,k} - \yQU \leq M_5 (1-\uBinary)\label{auxiliaryQu}\\
&0 \leq \yBetaU \leq M_{6_{i,j,m}} \uBinary\\
&0 \leq \beta_{t,k} - \yBetaU \leq M_{6_{i,j,m}} (1-\uBinary)\label{auxiliaryBetau}
\end{align}
where constraints \eqref{auxiliaryStart}-\eqref{auxiliaryMuMi} are defined $\forallt$, $\forallij$, $\forallm$, constraints \eqref{auxiliaryWii}-\eqref{auxiliaryLambdaQu} are defined $\forallt$, $\forall (i,j) \in \setLines \cup \tilde{\mathcal{L}}$, $\forallm$ with $\tilde{\mathcal{L}} = \{ (j,i): (i,j) \in \setLines\}$, constraints \eqref{auxiliaryPu}-\eqref{auxiliaryBetau} are defined $\forallt$, $\forallij$, $\forallm$,  $\forall k \in \setNodesPlus$. Moreover, $M_1$ is set equal to $3000$~p/kWh, which represents the maximum bid price currently allowed in the European wholesale markets, whereas: 
\begin{align}
&M_2=\sum_{k \in \setGenNode} \genMin - \sum_{k \in \setDemNode} \demMax - \demFixed\\ &M_3=\sum_{k \in \setGenNode} \genMax - \sum_{k \in \setDemNode} \demMin - \demFixed\\
&M_4=\sum_{k \in \setGenNode} \genReactiveMin - \sum_{k \in \setDemNode} \demReactiveMax\\
&M_5=\sum_{k \in \setGenNode} \genReactiveMax - \sum_{k \in \setDemNode} \demReactiveMin\\
&M_{6_{i,j,m}}=M_1\dfrac{\aRealAdmittance^2 + \bImagAdmittance^2}{\min(\aRealAdmittance, \bImagAdmittance)}
\end{align}

\section{Proof of \eqref{lemma1}}\label{sec:Appendix:proof}

By using the definition of $\activeInj$ in \eqref{lowerActiveInjDef} we have:
\begin{align}\label{proof5}
\nodalPrice\activeInj = \nodalPrice (\sum_{k \in \setGenNode} \gen - \sum_{k \in \setDemNode}\dem - \demFixed) = \sum_{k \in \setGenNode} \nodalPrice\gen - \sum_{k \in \setDemNode}\nodalPrice\dem - \nodalPrice\demFixed
\end{align}

The term $\nodalPrice\gen$ can be rewritten as follow by using \eqref{dualGen}:
\begin{align}
\nodalPrice\gen = &(\phiActiveGenMax - \phiActiveGenMin + \rhoUp - \rhoDown + \priceGen) \gen = \notag\\
&\phiActiveGenMax\gen - \phiActiveGenMin\gen + \rhoUp\gen - \rhoDown\gen + \priceGen\gen
\end{align}

The strong duality condition \eqref{strongDuality} ensures that all complementary slackness conditions hold. Therefore, by using the complementary slackness associated with \eqref{lowerGenActiveMax} we obtain:

\begin{align}
&(\gen - \genMax)\phiActiveGenMax = 0 && \iff && \gen \phiActiveGenMax = \genMax\phiActiveGenMax \label{proof1}\\
&(\genMin - \gen)\phiActiveGenMin = 0 && \iff && \gen \phiActiveGenMin = \genMin\phiActiveGenMin \label{proof2}
\end{align}

Therefore, by using \eqref{proof1}-\eqref{proof2}, and the dual conditions \eqref{dualReserveUp}-\eqref{dualReserveDown}, the following linear relation can be obtained:
\begin{align}\label{proof3}
\nodalPrice\gen = \phiActiveGenMax\genMax - \phiActiveGenMin\genMin + \priceReserveUp\sensitivityInjection\gen - \priceReserveDown\sensitivityInjection\gen + \priceGen\gen
\end{align}

Similarly, by using \eqref{dualDem}, the complementary slackness condition of  \eqref{lowerDemActiveMax}, and \eqref{dualReserveUp}-\eqref{dualReserveDown}, the following condition can be obtained:
\begin{align}\label{proof4}
\nodalPrice\dem = - \phiActiveDemMax\demMax + \phiActiveDemMin\demMin + \priceReserveUp\sensitivityInjection\dem - \priceReserveDown\sensitivityInjection\dem + \priceDem\dem
\end{align}

By substituting \eqref{proof3} and \eqref{proof4} in \eqref{proof5}, the relation stated in equivalence \eqref{lemma1} is obtained.

\section{Final MIQCP model}\label{sec:Appendix:finalMIQCP}

This section reports the final mixed-integer quadratically constrained program (where the bilinear terms in \eqref{strongDuality} have to be replaced with the respective auxiliary variables defined in \ref{sec:Appendix:auxiliaryVariables}, as detailed in Section \ref{sec:refomulationMIQCP}).

\begin{align}
\max_{} \quad& \sum_{t \in \setTime}\sum_{n \in \setNodesPlus}\Big( \sum_{k \in \setDemNode} \priceDem \dem - \sum_{k \in \setGenNode} \priceGen \gen \Big) 
- \sum_{t \in \setTime} \priceActivePowerZero \activeInjZero
- \sum_{t \in \setTime} \priceReactivePowerZero \reactiveInjZero
\notag\\
&
+ \sum_{t \in \setTime}\Big(\priceReserveUp \sum_{n \in \setNodesPlus} \sensitivityInjection \reserveUp
+ \priceReserveDown\sum_{n \in \setNodesPlus} \sensitivityInjection \reserveDown\Big) - \tariff \sum_{t \in \setTime}\sum_{n \in \setNodesPlus} \Big(\demFixed + \sum_{k \in \setDemNode}\demMax + \sum_{k \in \setGenNode} \genMax \Big)
\notag\\
&- \sum_{(i,j) \in \setLines}\sum_{m \in \setLumpy} \uBinary (\Kfix + \Kvar\Flumpy)\\
s.t.\notag\\
&\eqref{upperUniquePrice}\\
&\sum_{t \in \setTime} \sum_{n \in \setNodesPlus} \Big(\sum_{k \in \setGenNode} (-\phiActiveGenMax\genMax + \phiActiveGenMin\genMin - \priceReserveUp\sensitivityInjection\gen + \priceReserveDown\sensitivityInjection\gen - \priceGen\gen)\notag\\
&+ \sum_{k \in \setDemNode} (-\phiActiveDemMax\demMax + \phiActiveDemMin\demMin + \priceReserveUp\sensitivityInjection\dem - \priceReserveDown\sensitivityInjection\dem + \priceDem\dem) + \nodalPrice\demFixed\Big)  - \sum_{t \in \setTime}  \priceActivePowerZero\activeInjZero \notag\\
&+ \tariff \sum_{t \in \setTime}\sum_{n \in \setNodesPlus} \Big(\demFixed + \sum_{k \in \setDemNode}\demMax + \sum_{k \in \setGenNode} \genMax \Big) \geq \Kop + \sum_{j \in \setLumpy} \uBinary (\Kfix + \Kvar\Flumpy)\\
&\eqref{budget_constraint}-\eqref{upperVarDomain}\\
&\eqref{lowerActiveInjDef}-\eqref{lowerReactiveInjDef}
\end{align}
\begin{align}
&\activeInj = \sum_{(i,j) \in \setLines: i=n} \sum_{m \in \setLumpy}  \big( \yWiiu \aRealAdmittance - \yWijPu \aRealAdmittance + \yWijQu \bImagAdmittance \big)
+ \notag\\
&\qquad\sum_{(i,j) \in \setLines: j=n} \sum_{m \in \setLumpy}  \big( \yWjju \aRealAdmittance - \wjiP\uBinary \aRealAdmittance + \wjiQ\uBinary \bImagAdmittance \big) && \forall t \in \setTime, \forall n \in \setNodes  \\
&\reactiveInj = \sum_{(i,j) \in \setLines: i=n} \sum_{m \in \setLumpy}  \big( \yWiiu \bImagAdmittance - \yWijPu \bImagAdmittance - \yWijQu \aRealAdmittance \big) 
+ \notag\\
&\qquad\quad\sum_{(i,j) \in \setLines: j=n} \sum_{m \in \setLumpy}  \big( \yWjju \bImagAdmittance - \yWjiPu \bImagAdmittance - \yWijQu \aRealAdmittance \big) && \forall t \in \setTime, \forall n \in \setNodes \\
&\sum_{m \in \setLumpy}  \big(\yWiiu \aRealAdmittance - \yWijPu \aRealAdmittance + \yWijQu \bImagAdmittance \big)\leq \flowMax + \sum_{m \in \setLumpy} \uBinary \Flumpy &&
\forall t \in \setTime, \forall (i,j) \in \setLines && \\
&\sum_{m \in \setLumpy}  \big(\yWjju \aRealAdmittance - \yWjiPu \aRealAdmittance + \yWjiQu \bImagAdmittance \big) \leq \flowMin + \sum_{m \in \setLumpy} \uBinary \Flumpy && \forall t \in \setTime, \forall (i,j) \in \setLines\\
&W^{ii}_{t,0} + 2 \sum_{(i,j) \in \mathcal{P}_n} \sum_{k: i \in \mathcal{P}_k} \sum_{m \in \setLumpy} \big(\dfrac{\aRealAdmittance \yPU}{\aRealAdmittance^2 + \bImagAdmittance^2}  + \dfrac{\bImagAdmittance \yQU}{\aRealAdmittance^2 + \bImagAdmittance^2} \big) \leq (v^{max}_{t,n})^2 && \forallt, \forallnplus\\
&\eqref{lowervoltagemax}-\eqref{lowerSOCP}\\
&\eqref{dualDem}-\eqref{dualReactiveInjZero}
\end{align}
\begin{align}
&\nodalPrice + \lambdaPowerActive + 2 \sum_{(i,j) \in \mathcal{P}_n} \sum_{k: i \in \mathcal{P}_k} \sum_{m \in \setLumpy} \dfrac{\aRealAdmittance \yBetaU }{\aRealAdmittance^2 + \bImagAdmittance^2}  = 0 && \forall t \in \setTime, \forallnplus\\
&\nodalPriceRective + \lambdaPowerReactive + 2 \sum_{(i,j) \in \mathcal{P}_n} \sum_{k: i \in \mathcal{P}_k} \sum_{m \in \setLumpy}  \dfrac{\bImagAdmittance \yBetaU }{\aRealAdmittance^2 + \bImagAdmittance^2} = 0 && \forall t \in \setTime, \forallnplus \\
&\eqref{dualReserveUp}-\eqref{dualReserveDown}\\
&\sum_{m \in \setLumpy} \Big(\aRealAdmittance\ylambdaPu + \bImagAdmittance \ylambdaQu - \aRealAdmittance \ymuMaxU \Big) + \epsilonP + \etaOne = 0  && \forallt, \forallij \\
&\sum_{m \in \setLumpy} \Big(\aRealAdmittance\ylambdaPuJ + \bImagAdmittance \ylambdaQuJ - \aRealAdmittance \ymuMinU \Big) - \epsilonP  = 0  && \forallt, \forallij \\
&\sum_{m \in \setLumpy} \Big(-\bImagAdmittance \ylambdaPu + \aRealAdmittance \ylambdaQu + \bImagAdmittance \ymuMaxU \Big) + \epsilonQ + \etaTwo  = 0  && \forallt, \forallij \\
&\sum_{m \in \setLumpy} \Big(-\bImagAdmittance \ylambdaPuJ + \aRealAdmittance \ylambdaQuJ + \bImagAdmittance
\ymuMinU \Big) + \epsilonQ  = 0  && \forallt, \forallij 
\end{align}
\begin{align}
&\sum_{(i,j) \in \setLines: i=n} \Big( \sum_{m \in \setLumpy} \big(- \aRealAdmittance \ylambdaPu - \bImagAdmittance \ylambdaQu + \ymuMaxU \aRealAdmittance \big)
+ \dfrac{\etaThree - \gammaSOCP}{2}\Big) \notag\\ 
&+\sum_{(i,j) \in \setLines: j=n} \Big( \sum_{m \in \setLumpy} \big(- \aRealAdmittance \ylambdaPuJ - \bImagAdmittance \ylambdaQuJ + \ymuMinU \aRealAdmittance\big) - \dfrac{\etaThree + \gammaSOCP}{2}\Big) \notag\\
&+ \chiMax - \chiMin = 0 && \forallt,\forallnplus \\
&\eqref{dualSOCP}\\
&\eqref{strongDuality}\\
&\eqref{auxiliaryStart}-\eqref{auxiliaryBetau}
\end{align}

\section{Stochastic model}\label{sec:stoch}

\add{The proposed approach can be extended to consider uncertainty (e.g. in realizations of loads, bid prices, and available capacities of generators) by recasting the models reported in Section \ref{sec:subsec:upper_level} and Section \ref{sec:subsec:lowerLevel} as stochastic optimization problems through a scenario-based approach \cite{conejo2010decision}, detailed as follows. Each scenario is indexed by $\omega \in \mathcal{W}$, and has a probability of realization equal to $\pi_\omega$. Similarly, variables and parameters are indexed with the additional subscript $\omega$ to refer to each different scenario. For example, variables $d^p_{t,n,k,w}$ represent the active power for flexible consumer $k$ in node $n$ at time $t$ within the scenario $\omega$. By using this notation, the upper level problem in Section \ref{sec:subsec:upper_level} can be recast as a stochastic model as follow:
\begin{subequations}
	\begin{align}
	\max_{ \tariff, \uBinary} \quad \sum_{\omega \in \mathcal{W}} \pi_\omega & \Bigg( \sum_{t \in \setTime}\sum_{n \in \setNodesPlus}\Big( \sum_{k \in \setDemNode}  c^d_{t,n,k,\omega} d^p_{t,n,k,\omega} - \sum_{k \in \setGenNode} c^g_{t,n,k,\omega} g^p_{t,n,k,\omega} \Big) \Delta_t
	- \sum_{t \in \setTime} \Big(c^p_{t,0,\omega} p_{t,0,\omega} + c^q_{t,0,\omega} q_{t,0,\omega} \Big)\Delta_t
	\notag\\
	&+ \sum_{t \in \setTime}\Big(c^{up}_{t,0,\omega} \sum_{n \in \setNodesPlus} \sensitivityInjection r^{up}_{t,n,\omega}
	+ c^{down}_{t,0,\omega}\sum_{n \in \setNodesPlus} \sensitivityInjection r^{down}_{t,n,\omega}\Big)\Delta_t
	\notag\\
	&- \tariff \sum_{t \in \setTime}\sum_{n \in \setNodesPlus} \Big(D_{t,n,\omega} + \sum_{k \in \setDemNode}d^{p,max}_{t,n,k,\omega} + \sum_{k \in \setGenNode} g^{p,max}_{t,n,k,\omega} \Big)\Delta_t \Bigg)
	\notag\\
	&\hspace{-3em}
	- \sum_{(i,j) \in \setLines}\sum_{m \in \setLumpy} \uBinary (\Kfix + \Kvar\Flumpy)\label{stochUpper}\\
	s.t.\notag\\
	&- \sum_{\omega \in \mathcal{W}} \sum_{t \in \setTime} \pi_\omega \Big(\sum_{n \in \setNodes} \pi^p_{t,n,\omega} p_{t,n,\omega} 
	+ c^q_{t,0,\omega} q_{t,0,\omega}\Big) \Delta_t\notag\\
	&+ \tariff \sum_{\omega \in \mathcal{W}} \sum_{t \in \setTime}\sum_{n \in \setNodesPlus} \pi_\omega \Big(D_{t,n,\omega} + \sum_{k \in \setDemNode}d^{p,max}_{t,n,k,\omega} + \sum_{k \in \setGenNode} g^{p,max}_{t,n,k,\omega} \Big)\Delta_t \geq \notag\\
	&\Kop + \sum_{(i,j) \in \setLines}\sum_{m \in \setLumpy} \uBinary (\Kfix + \Kvar\Flumpy)\label{stochRevAd}\\
	&\eqref{budget_constraint}-\eqref{upperVarDomain}
	\end{align}
\end{subequations}
The main difference between the original objective function \eqref{upperObj} and \eqref{stochUpper} is that the latter maximizes the \textit{expected value} of the social welfare, while considering all scenarios, with $\omega \in \mathcal{W}$. Similarly, the left-hand side in \eqref{stochRevAd} represents the \textit{expected} revenues for the distribution network operator. Notice further that one instance of the lower level problem described in Section \ref{sec:subsec:lowerLevel} must be solved for each scenario $\omega$.}

\end{document}